\documentclass[twocolumn,letterpaper,aps,prc,longbibliography,superscriptaddress,showpacs,nofootinbib,floatfix]{revtex4-1}

\usepackage{graphicx}
\usepackage{dcolumn}
\usepackage{bm}
\usepackage{xspace}	
\usepackage{xcolor}
\usepackage{amsmath}
\usepackage{verbatim}
\usepackage{multirow}

\newcommand{\pt}{\mbox{$p_T$}\xspace}

\newcommand{\sqs}{\mbox{$\sqrt{s}$}\xspace}
\newcommand{\sqsn}{\mbox{$\sqrt{s_{_{NN}}}$}\xspace}
\newcommand{\dau}{\mbox{$d$$+$Au}\xspace}

\newcommand{\pau}{\mbox{$p$$+$Au}\xspace}

\newcommand{\hau}{\mbox{$^3$He$+$Au}\xspace}
\newcommand{\pp}{\mbox{$p$$+$$p$}\xspace}
\newcommand{\ppb}{\mbox{$p$$+$Pb}\xspace}

\newcommand{\bbceta}{\mbox{$3.0<|\eta|<3.9$}\xspace}

\newcommand{\supersonic}{{\sc supersonic}\xspace}

\begin{document}

\title{Measurements of mass-dependent azimuthal anisotropy in central
$p$$+$Au, $d$$+$Au, and $^3$He$+$Au collisions at $\sqrt{s_{_{NN}}}=200$ GeV}

\newcommand{\abilene}{Abilene Christian University, Abilene, Texas 79699, USA}
\newcommand{\augie}{Department of Physics, Augustana University, Sioux Falls, South Dakota 57197, USA}
\newcommand{\banaras}{Department of Physics, Banaras Hindu University, Varanasi 221005, India}
\newcommand{\barc}{Bhabha Atomic Research Centre, Bombay 400 085, India}
\newcommand{\baruch}{Baruch College, City University of New York, New York, New York, 10010 USA}
\newcommand{\bnlcoll}{Collider-Accelerator Department, Brookhaven National Laboratory, Upton, New York 11973-5000, USA}
\newcommand{\bnlphys}{Physics Department, Brookhaven National Laboratory, Upton, New York 11973-5000, USA}
\newcommand{\caucr}{University of California-Riverside, Riverside, California 92521, USA}
\newcommand{\charlesczech}{Charles University, Ovocn\'{y} trh 5, Praha 1, 116 36, Prague, Czech Republic}
\newcommand{\chonbuk}{Chonbuk National University, Jeonju, 561-756, Korea}
\newcommand{\ciae}{Science and Technology on Nuclear Data Laboratory, China Institute of Atomic Energy, Beijing 102413, People's Republic of China}
\newcommand{\cns}{Center for Nuclear Study, Graduate School of Science, University of Tokyo, 7-3-1 Hongo, Bunkyo, Tokyo 113-0033, Japan}
\newcommand{\colorado}{University of Colorado, Boulder, Colorado 80309, USA}
\newcommand{\columbia}{Columbia University, New York, New York 10027 and Nevis Laboratories, Irvington, New York 10533, USA}
\newcommand{\czechtech}{Czech Technical University, Zikova 4, 166 36 Prague 6, Czech Republic}
\newcommand{\debrecen}{Debrecen University, H-4010 Debrecen, Egyetem t{\'e}r 1, Hungary}
\newcommand{\elte}{ELTE, E{\"o}tv{\"o}s Lor{\'a}nd University, H-1117 Budapest, P{\'a}zm{\'a}ny P.~s.~1/A, Hungary}
\newcommand{\eszterhazy}{Eszterh\'azy K\'aroly University, K\'aroly R\'obert Campus, H-3200 Gy\"ongy\"os, M\'atrai \'ut 36, Hungary}
\newcommand{\ewha}{Ewha Womans University, Seoul 120-750, Korea}
\newcommand{\fsu}{Florida State University, Tallahassee, Florida 32306, USA}
\newcommand{\gsu}{Georgia State University, Atlanta, Georgia 30303, USA}
\newcommand{\hiroshima}{Hiroshima University, Kagamiyama, Higashi-Hiroshima 739-8526, Japan}
\newcommand{\howard}{Department of Physics and Astronomy, Howard University, Washington, DC 20059, USA}
\newcommand{\ihepprot}{IHEP Protvino, State Research Center of Russian Federation, Institute for High Energy Physics, Protvino, 142281, Russia}
\newcommand{\illuiuc}{University of Illinois at Urbana-Champaign, Urbana, Illinois 61801, USA}
\newcommand{\inrras}{Institute for Nuclear Research of the Russian Academy of Sciences, prospekt 60-letiya Oktyabrya 7a, Moscow 117312, Russia}
\newcommand{\instpasczech}{Institute of Physics, Academy of Sciences of the Czech Republic, Na Slovance 2, 182 21 Prague 8, Czech Republic}
\newcommand{\isu}{Iowa State University, Ames, Iowa 50011, USA}
\newcommand{\jaea}{Advanced Science Research Center, Japan Atomic Energy Agency, 2-4 Shirakata Shirane, Tokai-mura, Naka-gun, Ibaraki-ken 319-1195, Japan}
\newcommand{\jyvaskyla}{Helsinki Institute of Physics and University of Jyv{\"a}skyl{\"a}, P.O.Box 35, FI-40014 Jyv{\"a}skyl{\"a}, Finland}
\newcommand{\kek}{KEK, High Energy Accelerator Research Organization, Tsukuba, Ibaraki 305-0801, Japan}
\newcommand{\korea}{Korea University, Seoul, 136-701, Korea}
\newcommand{\kurchatov}{National Research Center ``Kurchatov Institute", Moscow, 123098 Russia}
\newcommand{\kyoto}{Kyoto University, Kyoto 606-8502, Japan}
\newcommand{\lawllnl}{Lawrence Livermore National Laboratory, Livermore, California 94550, USA}
\newcommand{\losalamos}{Los Alamos National Laboratory, Los Alamos, New Mexico 87545, USA}
\newcommand{\lund}{Department of Physics, Lund University, Box 118, SE-221 00 Lund, Sweden}
\newcommand{\lyon}{IPNL, CNRS/IN2P3, Univ Lyon, Université Lyon 1, F-69622, Villeurbanne, France}
\newcommand{\maryland}{University of Maryland, College Park, Maryland 20742, USA}
\newcommand{\mass}{Department of Physics, University of Massachusetts, Amherst, Massachusetts 01003-9337, USA}
\newcommand{\michigan}{Department of Physics, University of Michigan, Ann Arbor, Michigan 48109-1040, USA}
\newcommand{\muhlenberg}{Muhlenberg College, Allentown, Pennsylvania 18104-5586, USA}
\newcommand{\nara}{Nara Women's University, Kita-uoya Nishi-machi Nara 630-8506, Japan}
\newcommand{\natmephi}{National Research Nuclear University, MEPhI, Moscow Engineering Physics Institute, Moscow, 115409, Russia}
\newcommand{\newmex}{University of New Mexico, Albuquerque, New Mexico 87131, USA}
\newcommand{\nmsu}{New Mexico State University, Las Cruces, New Mexico 88003, USA}
\newcommand{\ohio}{Department of Physics and Astronomy, Ohio University, Athens, Ohio 45701, USA}
\newcommand{\ornl}{Oak Ridge National Laboratory, Oak Ridge, Tennessee 37831, USA}
\newcommand{\orsay}{IPN-Orsay, Univ.~Paris-Sud, CNRS/IN2P3, Universit\'e Paris-Saclay, BP1, F-91406, Orsay, France}
\newcommand{\peking}{Peking University, Beijing 100871, People's Republic of China}
\newcommand{\pnpi}{PNPI, Petersburg Nuclear Physics Institute, Gatchina, Leningrad region, 188300, Russia}
\newcommand{\riken}{RIKEN Nishina Center for Accelerator-Based Science, Wako, Saitama 351-0198, Japan}
\newcommand{\rikjrbrc}{RIKEN BNL Research Center, Brookhaven National Laboratory, Upton, New York 11973-5000, USA}
\newcommand{\rikkyo}{Physics Department, Rikkyo University, 3-34-1 Nishi-Ikebukuro, Toshima, Tokyo 171-8501, Japan}
\newcommand{\saispbstu}{Saint Petersburg State Polytechnic University, St.~Petersburg, 195251 Russia}
\newcommand{\seoulnat}{Department of Physics and Astronomy, Seoul National University, Seoul 151-742, Korea}
\newcommand{\stonybrkc}{Chemistry Department, Stony Brook University, SUNY, Stony Brook, New York 11794-3400, USA}
\newcommand{\stonycrkp}{Department of Physics and Astronomy, Stony Brook University, SUNY, Stony Brook, New York 11794-3800, USA}
\newcommand{\tenn}{University of Tennessee, Knoxville, Tennessee 37996, USA}
\newcommand{\titech}{Department of Physics, Tokyo Institute of Technology, Oh-okayama, Meguro, Tokyo 152-8551, Japan}
\newcommand{\tsukuba}{Tomonaga Center for the History of the Universe, University of Tsukuba, Tsukuba, Ibaraki 305, Japan}
\newcommand{\vandy}{Vanderbilt University, Nashville, Tennessee 37235, USA}
\newcommand{\weizmann}{Weizmann Institute, Rehovot 76100, Israel}
\newcommand{\wigner}{Institute for Particle and Nuclear Physics, Wigner Research Centre for Physics, Hungarian Academy of Sciences (Wigner RCP, RMKI) H-1525 Budapest 114, POBox 49, Budapest, Hungary}
\newcommand{\yonsei}{Yonsei University, IPAP, Seoul 120-749, Korea}
\newcommand{\zagreb}{Department of Physics, Faculty of Science, University of Zagreb, Bijeni\v{c}ka c.~32 HR-10002 Zagreb, Croatia}
\affiliation{\abilene}
\affiliation{\augie}
\affiliation{\banaras}
\affiliation{\barc}
\affiliation{\baruch}
\affiliation{\bnlcoll}
\affiliation{\bnlphys}
\affiliation{\caucr}
\affiliation{\charlesczech}
\affiliation{\chonbuk}
\affiliation{\ciae}
\affiliation{\cns}
\affiliation{\colorado}
\affiliation{\columbia}
\affiliation{\czechtech}
\affiliation{\debrecen}
\affiliation{\elte}
\affiliation{\eszterhazy}
\affiliation{\ewha}
\affiliation{\fsu}
\affiliation{\gsu}
\affiliation{\hiroshima}
\affiliation{\howard}
\affiliation{\ihepprot}
\affiliation{\illuiuc}
\affiliation{\inrras}
\affiliation{\instpasczech}
\affiliation{\isu}
\affiliation{\jaea}
\affiliation{\jyvaskyla}
\affiliation{\kek}
\affiliation{\korea}
\affiliation{\kurchatov}
\affiliation{\kyoto}
\affiliation{\lawllnl}
\affiliation{\losalamos}
\affiliation{\lund}
\affiliation{\lyon}
\affiliation{\maryland}
\affiliation{\mass}
\affiliation{\michigan}
\affiliation{\muhlenberg}
\affiliation{\nara}
\affiliation{\natmephi}
\affiliation{\newmex}
\affiliation{\nmsu}
\affiliation{\ohio}
\affiliation{\ornl}
\affiliation{\orsay}
\affiliation{\peking}
\affiliation{\pnpi}
\affiliation{\riken}
\affiliation{\rikjrbrc}
\affiliation{\rikkyo}
\affiliation{\saispbstu}
\affiliation{\seoulnat}
\affiliation{\stonybrkc}
\affiliation{\stonycrkp}
\affiliation{\tenn}
\affiliation{\titech}
\affiliation{\tsukuba}
\affiliation{\vandy}
\affiliation{\weizmann}
\affiliation{\wigner}
\affiliation{\yonsei}
\affiliation{\zagreb}
\author{A.~Adare} \affiliation{\colorado} 
\author{C.~Aidala} \affiliation{\michigan} 
\author{N.N.~Ajitanand} \altaffiliation{Deceased} \affiliation{\stonybrkc} 
\author{Y.~Akiba} \email[PHENIX Spokesperson: ]{akiba@rcf.rhic.bnl.gov} \affiliation{\riken} \affiliation{\rikjrbrc} 
\author{M.~Alfred} \affiliation{\howard} 
\author{V.~Andrieux} \affiliation{\michigan} 
\author{N.~Apadula} \affiliation{\isu} \affiliation{\stonycrkp} 
\author{H.~Asano} \affiliation{\kyoto} \affiliation{\riken} 
\author{B.~Azmoun} \affiliation{\bnlphys} 
\author{V.~Babintsev} \affiliation{\ihepprot} 
% ...
\collaboration{PHENIX Collaboration} \noaffiliation

\date{\today}


\begin{abstract}


We present measurements of the transverse-momentum dependence of 
elliptic flow $v_2$ for identified pions and (anti)protons at 
midrapidity ($|\eta|<0.35$), in 0\%--5\% central $p$$+$Au and 
$^3$He$+$Au collisions at $\sqrt{s_{_{NN}}}=200$ GeV.  When taken 
together with previously published measurements in $d$$+$Au collisions 
at $\sqrt{s_{_{NN}}}=200$ GeV, the results cover a broad range of 
small-collision-system multiplicities and intrinsic initial geometries. We 
observe a clear mass-dependent splitting of $v_2(p_{T})$ in $d$$+$Au and 
$^3$He$+$Au collisions, just as in large nucleus-nucleus ($A$$+$$A$) 
collisions, and a smaller splitting in $p$$+$Au collisions.  Both 
hydrodynamic and transport model calculations successfully describe the 
data at low $p_T$ ($<1.5$~GeV/$c$), but fail to describe various features 
at higher $p_T$. In all systems, the $v_2$ values follow an 
approximate quark-number scaling as a function of the hadron transverse 
kinetic energy per constituent quark($KE_T/n_q$), which was also 
seen previously in $A$$+$$A$ collisions.

\end{abstract}

\maketitle

\section{Introduction}

Recent years have seen a paradigm shift in our understanding of the 
minimum conditions required for the production of the quark-gluon plasma 
(QGP). In large nucleus-nucleus ($A$$+$$A$) collisions, signals of collective 
behavior---such as the azimuthal momentum anisotropy of final-state 
particles relative to the event plane---have been successfully 
understood in the context of nearly-inviscid hydrodynamic calculations, 
thus establishing the notion of a strongly interacting, nearly-perfect 
fluid being formed in this class of collisions~\cite{Heinz:2013th}.

However, the discovery of the same azimuthal anisotropy signals in a 
variety of small collision systems (i.e, $p,d,^{3}\text{He}$+Au at \sqsn 
= 200 GeV~\cite{PhysRevLett.111.212301,Adare:2014keg,Adare:2015ctn}; 
\ppb at \sqsn = 5.02 TeV; \pp at \sqs = 2.76, 5.02, and 13 
TeV~\cite{alice_long_2013,atlas_observation_2012,cms_observation_2012,Khachatryan:2015lva,Aad:2015gqa,Khachatryan:2010gv,Khachatryan:2016txc}; 
and an earlier observation of long-range two-particle correlations in 
\pp collisions at \sqs = 7 TeV~\cite{Khachatryan:2010gv}) pose a 
challenge. It was believed that the system size in this class of 
collisions is too small to create any significant amount of hot nuclear 
matter, which in any case would be very short lived.  There are also 
alternative explanations for these anisotropy signals based on momentum 
space domains and color recombination, such 
as~\cite{Dusling:2012iga,Ortiz:2013yxa}, although they lack quantitative 
predictions for small-system observables at the Relativistic Heavy Ion 
Collider (RHIC).  Therefore, in small collision systems, the 
identification of collective behavior with the hydrodynamic expansion of 
any potential QGP requires further scrutiny.

Measurements of elliptic and triangular flow ($v_2$ , $v_3$) at RHIC in 
\hau collisions, as well as of $v_2$ in \dau and \pau collisions, 
demonstrated that the observed collective response in small collision 
systems is directly correlated with the event 
geometry~\cite{Adare:2015ctn,Adare:2014keg,PhysRevC.95.034910}, just as 
in $A$$+$$A$ collisions where the geometric configuration of the 
overlapping nuclei determines the pressure gradients that drive the 
expansion of the resulting QGP. Viscous hydrodynamic calculations 
successfully describe the measurements in the geometry-controlled 
experiments at 
RHIC~\cite{nagle_exploiting_2013,Bozek:2015qpa,Schenke:2014gaa,Shen:2016zpp}, 
as well as those made at the Large Hadron Collider (LHC) in \ppb, and 
even in \pp collisions~\cite{Weller:2017tsr}. The success of 
hydrodynamics in describing small-system collectivity over such a wide 
range of energies and for a variety of systems is taken as evidence for 
the claim that the QGP is formed in these collisions and through its 
expansion translates initial geometry into final-state momentum 
anisotropy.

If collectivity in small systems can indeed be understood as arising 
from the expansion of QGP droplets along pressure gradients determined 
by geometry, there should necessarily be a mass ordering of $v_2(\pt )$ 
for identified final-state hadrons.  Strong radial expansion in the 
hydrodynamic evolution results in a shifting of the anisotropy pattern 
to higher \pt for higher mass hadrons due to a common velocity 
boost~\cite{Heinz:2013th}.  This fingerprint of hydrodynamic expansion 
on the $v_{2}(m,p_{T})$ is one of the key signatures of the nearly 
inviscid fluid nature of the QGP formed in $A$$+$$A$ collisions--- see for 
example ~\cite{Adler:2003kt}. Recently, such mass ordering has been 
observed in \dau collisions at RHIC~\cite{Adare:2014keg} and in \ppb 
collisions at the LHC~\cite{ABELEV:2013wsa,Khachatryan:2014jra}.

It is notable that a-multiphase-transport model (\textsc{ampt}), an 
instance of a broader family of kinetic transport 
models~\cite{lin_multiphase_2005}, also finds a mass ordering of 
$v_{2}(\pt)$ in both $A$$+$$A$ and small systems, despite having only a modest 
number of parton scatterings and thus nothing close to a radial velocity 
field as in hydrodynamics~\cite{Li:2016flp}.  Within \textsc{ampt} the 
mass ordering is found to arise from the hadronic rescattering phase, 
after all partons have coalesced into hadrons, incorporating the 
different inelastic cross sections for different 
hadrons~\cite{Li:2016flp}.  There are other approaches with 
fragmentation of saturated gluon states~\cite{Schenke:2016lrs} and with 
color strings followed by hydrodynamics~\cite{Werner:2013ipa} that 
achieve some degree of mass ordering, though currently lacking in any 
predictions for small systems at RHIC energies.

The present study completes the set of small-system projectile geometry 
results at top RHIC energy by providing $v_2$ measurements for pions and 
(anti)protons (henceforth referred to as ``protons'') in central \pau 
and \hau collisions at \sqsn = 200 GeV, and compares to $v_2$ 
measurements for pions and (anti)protons in central \dau collisions at 
the same energy~\cite{Adare:2014keg}. Detailed comparisons are then made 
with theory calculations from viscous hydrodynamics, as encoded in the 
\supersonic~\cite{Romatschke2015} and the i\textsc{ebe-vishnu} 
models~\cite{Shen:2016zpp}, and the kinetic transport model 
\textsc{ampt}.


\section{Methods}

The PHENIX collaboration has measured the $v_{2}(\pt)$ of identified 
particles in \pau, \dau and \hau collisions. We apply the same analysis 
procedure to all three systems in the same centrality class, to provide 
a controlled comparison from which to draw conclusions.

A complete description of the PHENIX detector and its subsystems can be 
found in ~\cite{Adcox2003469,Aidala:2013vna}. Charged particles are 
reconstructed with the two central arm spectrometers, comprising drift 
chambers (DC) and  multi-wire proportional pad chambers (PC). Each arm covers an acceptance of $|\eta|<0.35$ in pseudorapidity and $\pi/2$ in 
azimuth. Tracks in the drift chamber are matched to hits in the 
outer detectors. The distribution of differences between hits and projections is approximately Gaussian, with an additional underlying background caused by random associations. 
To suppress background from particle weak decays and photon conversions, tracks reconstructed with the DC and the first layer of PC are required to be matched to the third layer of PC within three $\sigma$ in the longitudinal and transverse planes, where \pt and charge sign dependent $\sigma$ values are determined from Gaussian fits to residual distributions between PC signals/clusters and the tracks extrapolated to the PC surface. Particle identification is performed using the TOF subsystem, which comprises 
two separate arms (east and west), constructed using scintillators~\cite{AIZAWA2003508} 
and multi-gap resistive plate chambers~\cite{Adare:2013esx}, 
and covers  $\pi/4$ and  $\pi/8$, respectively. The timing resolutions for the east and west TOF are 130 ps and 95 ps, respectively. 
Particle identification (PID) is based on the particle mass with \pt-dependent selections in mass-squared calculated using the particle momentum, time-of-flight and the path length. After track matching and PID selections, some residual background remains in the proton sample at low \pt ($< 1$~GeV/$c$).
In this \pt range, up to 15\% of the reconstructed protons are secondary particles that originate from  interactions of energetic particles produced in the collisions with  detector material, primarily the silicon vertex tracker (VTX), which  
covers the pseudorapidity range $|\eta| < 1.2$.  Detector simulations 
using \textsc{GEANT3} \cite{geant} indicate that the contamination in 
the proton sample is negligible for \pt$>1$~GeV/$c$, not present in the 
anti-proton distributions, and negligible in the charged pion sample at 
all $p_{T}$. To remove the background in the proton sample, the VTX 
detector~\cite{Adare:2015hla} is used in conjunction with the DC to 
select proton tracks with \pt$< 1$~GeV/$c$ based on their distance of 
closest approach (DCA) to the primary vertex in the x-y plane transverse 
to the beam direction. The tracks are required to be within two standard 
deviations of the mean value of the DCA distribution. This additional 
selection is not applied at higher \pt nor for particle species for 
which the secondary-particle contamination is negligible.  The pions and 
protons selected for the analysis are identified with purity of over 
98\% for \pt up to 3 GeV/$c$ in all collision systems.

The beam-beam counters (BBC) comprise two arrays of 64 quartz radiator 
\v{C}erenkov detectors, placed longitudinally $\pm$1.44 m away from the 
center of the interaction region (IR), covering \bbceta and 2$\pi$ in 
azimuth. The forward vertex detector (FVTX) is a silicon detector 
comprised of two identical end-cap assemblies symmetrically arranged in 
the longitudinal direction around the IR, covering the pseudorapidity 
acceptance $1.0 < |\eta| < 3.0$. Using hit clusters, it can detect 
charged particles with an efficiency greater than 95\%. The arms of the 
BBC and FVTX in the Au-going direction (i.e., $\eta < 0$) are designated 
as the \emph{south} arms and designated BBC-S and FVTX-S, respectively. 
We use the south arm of each of these detectors to determine the event 
plane of the collision. In addition, timing information from the BBC is 
used to determine the $z$-vertex of the collision. In this analysis, a 
$\pm$10 cm cut on the collision $z$-vertex is applied.

The \pau data set for this analysis, taken during the 2015 run at RHIC, 
comprises 0.84 billion minimum bias (MB) triggered events and 1.4 
billion high-multiplicity (HM) triggered events. The MB trigger is 
defined as a coincidence in the same event between both arms of the BBC 
detector~\cite{Allen2003549}, requiring that at least one 
photomultiplier tube (PMT) fire in each. This definition allows 
84$\pm$4\% of the total inelastic \pau cross section to be captured. The 
HM trigger is based on the MB trigger, but with the additional 
requirement of more than 35 photomultiplier tubes firing in the BBC-S. 
Events that satisfy this trigger condition correspond roughly to the 5\% 
most central event class. The use of this high-multiplicity trigger 
allows us to increase our central \pau event sample size by a factor of 
25. The \hau data set for this analysis was recorded during the 2014 run 
at RHIC, and comprises 1.6 billion MB events and 480 million HM 
events. The HM trigger used in \hau is also based on the MB trigger, but 
with the additional requirement of more than 48 photomultiplier tubes 
firing in the BBC-S. The \dau data set was recorded during the 2008 run, 
and comprises 1.56 billion MB events.

In this analysis, we select the 0\%--5\% most central events in all 
collision systems, where centrality classes are defined as percentiles 
of the total charged particle multiplicity as measured in the BBC-S, 
following the procedure presented in ~\cite{bbc}. We follow the 
identical analysis procedure that was previously used in \hau and \pau 
collisions~\cite{Adare:2015ctn,PhysRevC.95.034910} to measure $v_2$ for 
inclusive charged hadrons. Namely, we measure $v_2$ for final-state 
single hadrons at midrapidity with respect to the event 
plane~\cite{Poskanzer:1998yz} of the collision, as follows:

\begin{equation}
v_{2}(p_{T}) = \frac{\langle \cos 2(\phi_{\text{Particle}}(p_{T})-\Psi^{\text{FVTX-S}}_{2})\rangle}{\text{Res}(\Psi^{\text{FVTX-S}}_{2})}.
\end{equation}
The event-plane angle is determined by the event flow vector $Q_2$ measured in the Au-going direction where the particle multiplicity is higher. The Q-vectors are re-centered according to the standard procedure described in~\cite{Poskanzer:1998yz}.
The raw event plane angle is estimated by:
\begin{equation}
 \Psi_n^{\rm raw} = {\rm atan2}(Q_2^y,Q_2^x)/2,
\end{equation}
where $Q_2^x$ and $Q_2^y$ are the $x$ and $y$ projections of the flow vector. A standard flattening procedure described in~\cite{Poskanzer:1998yz} is applied to the $\Psi_2^{\text{raw}}$ distributions to remove detector acceptance effects. 
The second order event-plane angle $\Psi^{\text{FVTX-S}}_{2}$ is 
determined using the FVTX-S detector. Its resolution 
$\text{Res}(\Psi_{2})$ is evaluated using the standard three-subevent 
method~\cite{Poskanzer:1998yz}, correlating independent measurements 
made in the BBC-S, FVTX-S, and the central arms. The resolution of the 
event plane is found to be $\text{Res}(\Psi^{\text{FVTX-S, $p$$+$Au}}_{2})$ 
= 0.171 in \pau collisions, and $\text{Res}(\Psi^{\text{FVTX-S}, 
^{3}\text{He+Au}}_{2})$ = 0.274 in \hau collisions. If the event plane 
is instead measured using the BBC-S detector, we obtain a lower 
resolution $\text{Res}(\Psi^{\text{BBC-S, $p$$+$Au}}_{2})$ = 0.062 in \pau 
and $\text{Res}(\Psi^{\text{BBC-S,}^{3}\text{He+Au}}_{2})$ = 0.070 in 
\hau collisions. The event-plane resolution depends on the particle 
multiplicity registered in the detectors used for event-plane 
determination, which results in better resolution in FVTX-S than in 
BBC-S.

\section{Systematic Uncertainties}
We identify the following as the main sources of systematic uncertainty 
in the $v_2(p_{T})$ measurement:

\begin{enumerate}

\item \textit{Background tracks from weak decays, photon conversions, 
and misreconstructed tracks.} We estimate the magnitude of this 
uncertainty by narrowing the spatial matching windows of the tracks and 
the hits in the outermost layer of the PC, from 3$\sigma$ to 2$\sigma$ 
and comparing the resulting values of $v_2(\pt)$. The relative 
uncertainty in $v_2$ is 2\% in both \pau and \hau collisions.

\item \textit{Multiple collisions per bunch crossing}. Also referred to 
as event pile-up, these are observed to occur at an average rate of 8\% 
(4\%--5\%) in the centrality class of interest in \pau ( \hau) 
collisions.  We estimate the associated systematic uncertainty by 
analyzing low- and high-luminosity subsets of the data. The measured 
$v_2$ was found to decrease in events with higher pile-up rate, and an 
asymmetric systematic uncertainty of $^{+4}_{-0}\%$ was assigned in 
\pau, and $^{+5}_{-0}\%$ was assigned in \hau collisions.

\item \textit{Nonflow correlations from elementary processes.} There are 
many sources of correlations among particles which enhance the measured 
$v_2$, yet are unrelated to collective flow, such as momentum 
conservation.  We use a reference method previously employed in PHENIX 
analyses of small-system collectivity~\cite{PhysRevC.95.034910} to 
assign a $p_T$-dependent asymmetric uncertainty with a maximum value of 
$^{+0}_{-23}\%$ for the highest $p_T$ bin in \pau collisions. This can 
be compared to the corresponding values of 
$^{+0}_{-9}\%$~\cite{Adare:2014keg} and 
$^{+0}_{-7}\%$~\cite{Adare:2015ctn} in \dau and \hau collisions, 
respectively.  The nonflow effect has a larger relative contribution in 
\pau collisions due to the smaller multiplicity in this system.

\item \textit{Detector acceptance asymmetry.} In \pau collisions, there 
exists an asymmetry between the east ($\pi/2 < \phi < 3\pi/2$) and west 
($-\pi/2 < \phi < \pi/2$) acceptance of the detectors, originating from 
a 3.6 mrad offset between the beams at the collision point and the 
longitudinal axis of PHENIX.  This offset is necessary to compare to  
\pau collisions at the same momentum per nucleon.  We account for this 
effect by performing a counter-rotation on every central arm track and 
detector element in the FVTX and the BBC, taking care to restore their 
azimuthal anisotropy by re-weighting. There remains a small residual 
asymmetry after applying these corrections in \pau. Meanwhile in \hau 
collisions this beam angle is negligible and we assign a value of 5\% 
for this systematic uncertainty by taking the difference of $v_2$ when 
measured exclusively in the east or west arms in both \pau and \hau 
collisions.

\item \textit{Event plane measured with different detectors.} We observe 
the measured $v_2(p_T)$ to differ when using the event plane as 
determined using the BBC-S or the FVTX-S detectors. Despite the large 
difference in event-plane resolution in these two detectors, the 
differences in the measured $v_2(\pt)$ values are only of the order 3\% 
in \pau, and 5\% in \hau collisions, which demonstrates that the 
corrections for event-plane resolution are well understood.

\item \textit{Particle identification purity.} The effect of particle 
identification purity on the measured $v_2$ values is evaluated by 
varying the width of particle selection windows in the mass-squared vs 
\pt space from 2~$\sigma$ to 1.5~$\sigma$.  The uncertainty is found to 
be at most 2\% for both pions and protons in both collision systems.

\end{enumerate}
\label{s:sys}

\begin{table}[htbp]
\caption{\label{t:sys}Systematic uncertainties given as a percent of the 
$v_2$ measurement. Note that the nonflow contribution is \pt dependent 
and the quoted values corresponds to the highest measured \pt.
}
\begin{ruledtabular}      \begin{tabular}{cccccc}
&      Source& $p$$+$Au & $^3$He$+$Au & Type              &\\ 
\hline
&      Track Background &2\%& 2\% & A                     &\\ 
&      Event Pile-up    &$^{+4}_{-0}\%$&$^{+5}_{-0}\%$& B &\\
&      Nonflow    &$^{+0}_{-23}\%$&$^{+0}_{-7}\%$& B      &\\
&      Acceptance Asymmetry &5\%&5\%& C                   &\\  
&      Event-Plane Detectors & 3\% & 5\% &C               &\\
&      Particle Purity &2\% & 2\%& B                      &\\
    \end{tabular} \end{ruledtabular}
 \end{table}

Table~\ref{t:sys} summarizes all these systematic uncertainties, 
categorized by type:

\begin{itemize}
\item[A] point-to-point uncorrelated between $p_T$ bins,
\item[B] point-to-point correlated between $p_T$ bins,
\item[C] overall normalization uncertainty in which all data points are 
scaled by the same multiplicative factor.
\end{itemize}


\section{Results and discussion}

\begin{figure*}[htb]
\begin{minipage}{0.99\linewidth}
  \includegraphics[width=1.0\linewidth]{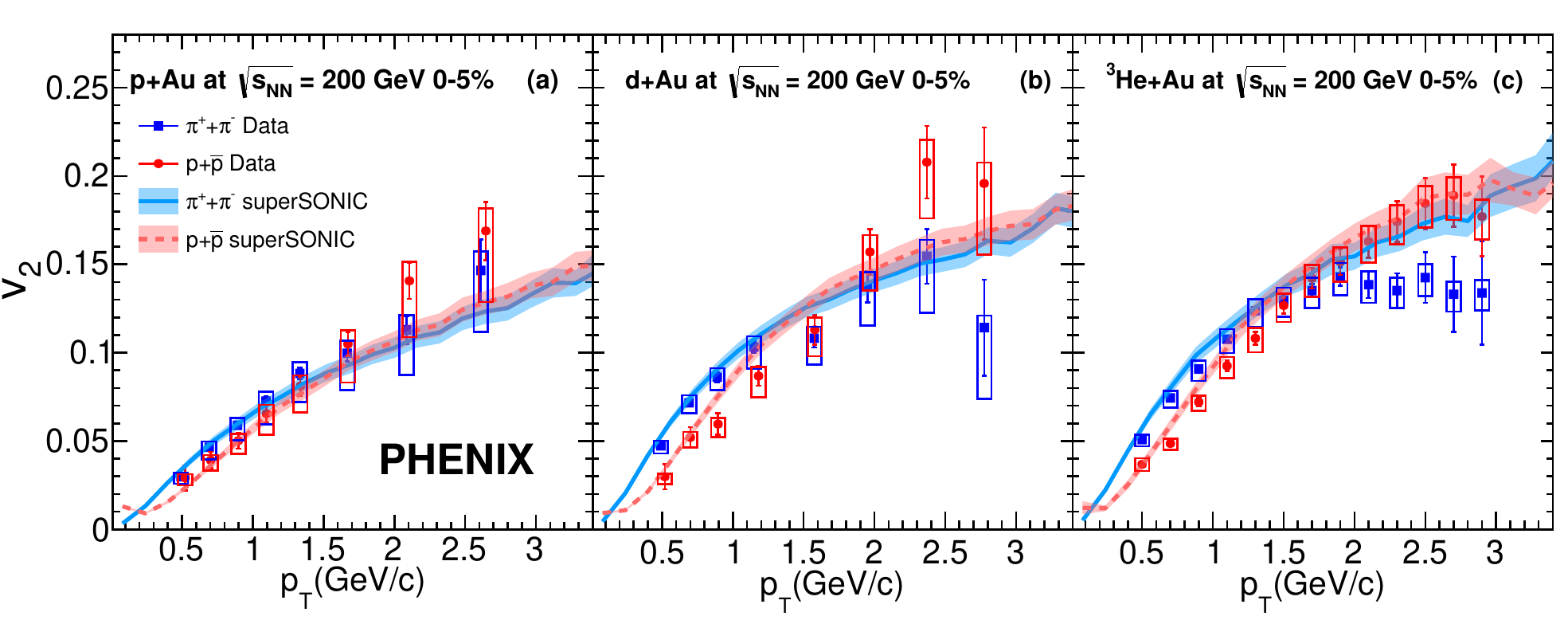}
\caption{Transverse momentum dependence of $v_2$ for identified pions 
and protons within $|\eta| <$ 0.35 in 0\%--5\% central \pau, 
\dau~\cite{Adare:2014keg}, and \hau collisions. The measurements are 
compared to hydrodynamic calculations using the \supersonic 
model~\cite{Habich:2014jna}, matched to the same multiplicity at 
midrapidity as the data. Note that the data points shown include nonflow 
contributions, whose estimated magnitude is accounted for in the 
asymmetric systematic uncertainties.
}
\label{fig:figure_whydro}
\end{minipage}
\begin{minipage}{0.99\linewidth}
  \includegraphics[width=1.0\linewidth]{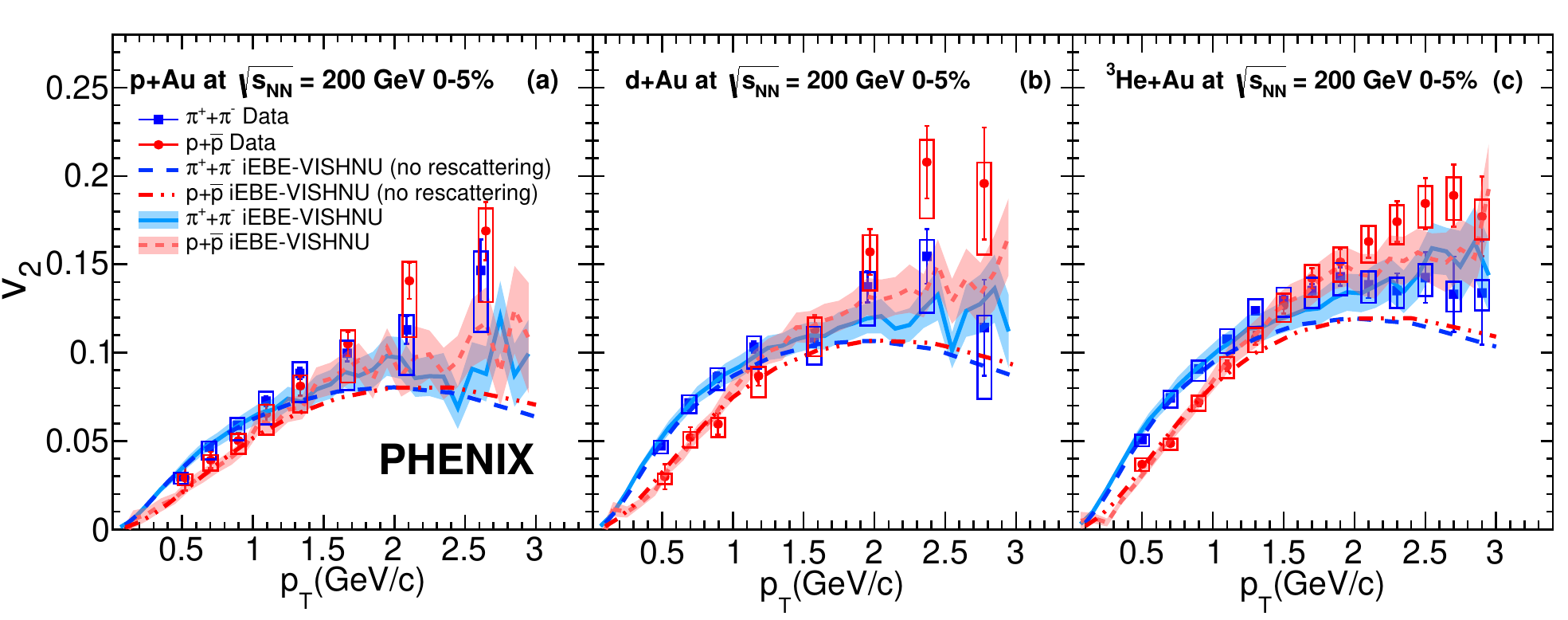}
\caption{Same as Fig.~\protect\ref{fig:figure_whydro}, but also shown 
are $v_2(p_T)$ calculations using the i\textsc{ebe-vishnu} hydrodynamic 
model~\cite{Shen:2016zpp}, illustrating the effect of hadronic 
rescattering on the mass-dependent $v_2$ values.
}
\label{fig:figure_whydro2}
\end{minipage}
\begin{minipage}{0.99\linewidth}
  \includegraphics[width=1.0\linewidth]{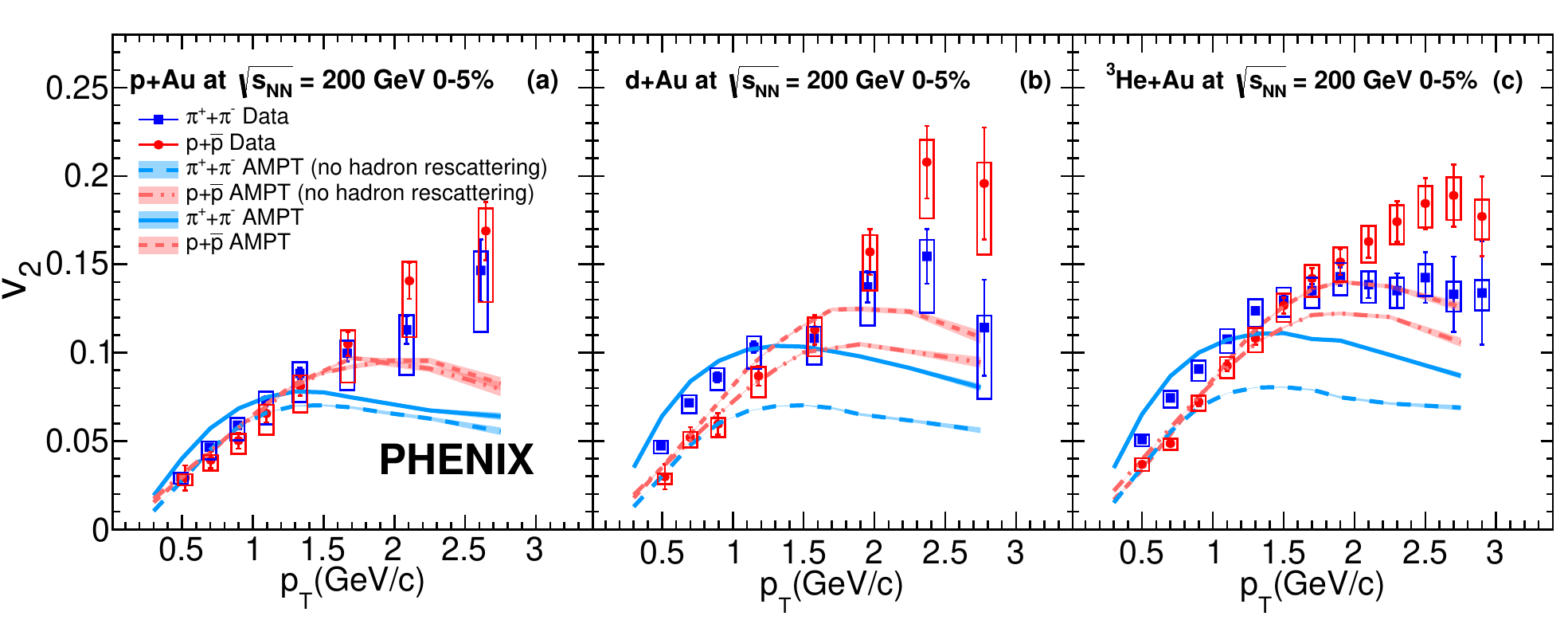}
\caption{Same as Fig.~\protect\ref{fig:figure_whydro}, but 
also shown are $v_2(p_T)$ transport model calculations using 
\textsc{ampt}~\cite{lin_multiphase_2005}.
}
\label{fig:figure_wampt}
\end{minipage}
\end{figure*}

Figure~\ref{fig:figure_whydro} shows $v_2(p_T)$ for identified pions and 
protons in 0\%-5\% central \pau, \dau ~\cite{Adare:2014keg}, and \hau 
collisions.  For both pions and protons the $v_{2}(\pt)$ values are 
higher in \dau and \hau collisions than in \pau collisions, as 
previously observed for inclusive charged 
hadrons~\cite{PhysRevC.95.034910}. These values follow the ordering of 
the initial geometric eccentricity $\varepsilon_{2}$(\pau) $<$ 
$\varepsilon_{2}$(\hau) $\approx$ $\varepsilon_{2}$(\dau).

In the \dau and \hau systems, there is a clear separation between the 
pion and proton $v_{2}$, with the pion $v_{2}$ being larger than the 
proton $v_{2}$ for $p_{T} \lesssim 1.5$~GeV/$c$ and this order being 
reversed at higher $p_{T}$. In the \pau system, the pion and proton 
$v_{2}(\pt)$ values show smaller overall splitting. The splitting 
pattern and the reversal of the mass ordering above $p_{T} \gtrsim 
1.5$~GeV/$c$ is qualitatively the same as has been observed in Au$+$Au 
collisions at \sqsn = 200~GeV~\cite{Adler:2003kt, PhysRevLett.87.182301}.

Figure~\ref{fig:figure_whydro} compares the measured $v_2(p_T)$ with 
hydrodynamic calculations using the \supersonic 
model~\cite{Habich:2014jna}. This model comprises standard Monte Carlo 
Glauber initial conditions followed by a viscous hydrodynamic expansion 
stage with $\eta/s=0.08$, Cooper-Frye hadronization at $T=$ 170 MeV, and 
a subsequent hadronic cascade code, B3D~\cite{PhysRevC.89.034917}. The 
\supersonic model additionally incorporates pre-equilibrium dynamics via 
a calculation in the context of the AdS/CFT 
correspondence~\cite{vanderSchee:2013pia,Chesler:2015wra,Romatschke:2013re}. 
These hydrodynamic calculations are matched to the measured charged 
particle density at midrapidity in the 0\%-5\% centrality class for \dau 
and \hau (i.e., $dN_{ch}/d\eta =$ 20.0 and 27.0, 
respectively~\cite{Adare:2015bua}). Because $dN_{ch}/d\eta$ has not yet 
been measured in \pau collisions, a value of $dN_{ch}/d\eta = 10.0$ was 
extrapolated for this system~\cite{Habich:2014jna}.

We observe that the hydrodynamic calculations agree with the data within 
uncertainties at low $p_T$, but fail to describe the reversal of the 
pion and proton $v_{2}$ ordering for $p_{T} > 1.5$~GeV/$c$. Viscous 
hydrodynamic calculations similarly describe Au$+$Au $v_{2}$ data at low 
$p_{T}$ but do not match the strong reverse ordering at higher $p_{T}$. 
For $p_{T} < 1.5$~GeV/$c$, the mass splitting increases in going from 
\pau to \dau and \hau as also seen in the data. Within the context of 
hydrodynamic calculations, this is due to the increased radial flow and 
consequently larger velocity boost when going from the smaller and lower 
multiplicity system to the larger and higher multiplicity systems.

In the case of ideal hydrodynamics, i.e. with zero viscosity, the 
$v_{2}$ values for all hadrons asymptotically approach each other at 
high $p_{T}$~\cite{Huovinen:2001cy}.  However, viscous effects and the 
incorporation of late stage hadronic rescattering have the effect of 
lowering the high $p_{T}$ $v_{2}$ values, more strongly so for pions. 
This can be seen in the \supersonic calculations. However, the predicted 
high $p_{T}$ splitting is much smaller than that seen in the \dau and 
\hau data.  It is in this high $p_{T}$ region in $A$$+$$A$ collisions that 
proposals of hadronization via recombination~\cite{Fries:2008hs} have 
been set forth to explain the $v_{2}$ splitting as well as the 
observation of enhanced baryon yields~\cite{Adler:2003kg,Adler:2003cb}.

\begin{figure*}[tbh]
  \includegraphics[width=1.0\linewidth]{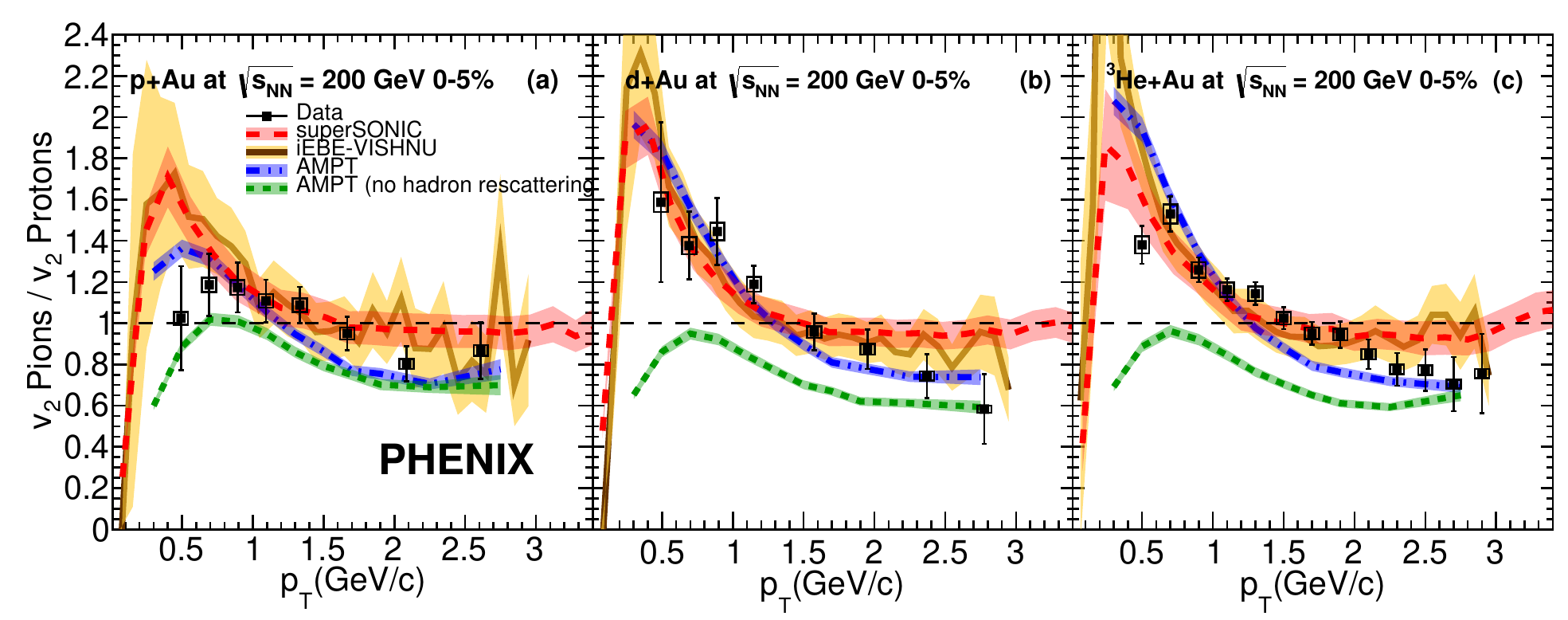}
  \caption{ Ratio of $v_2^{Pion}$ over $v_2^{Proton}$ in central 0\%-5\% 
(a) \pau, (b) \dau, and (c) \hau collisions at \sqsn = 200 GeV. 
Theoretical calculations from \supersonic and \textsc{ampt} are also shown. 
}
\label{fig:figure_ratio}
\end{figure*}

Figure~\ref{fig:figure_whydro2} shows results from another viscous 
hydrodynamic calculation, i\textsc{ebe-vishnu}~\cite{Shen:2016zpp}.  The 
calculation includes event-by-event fluctuating initial conditions via 
Monte Carlo Glauber simulation and then viscous hydrodynamics starting 
at $\tau_{0}$ = 0.6~fm/c.  The hydrodynamic evolution utilizes an 
$\eta/s = 0.08$ for RHIC energies and ends at $T = 155$~MeV.  After that 
point, hadronization occurs and hadronic rescattering is implemented 
using \textsc{urqmd}~3.4~\cite{Bass:1998ca,Bleicher:1999xi}.  The 
calculation results with viscous hydrodynamics followed by hadronic 
rescattering show good agreement with the experimental data for all 
three small systems.  Also shown are results with no hadronic 
rescattering that reveal almost no change in the $v_{2}$ for pions and 
protons for $p_{T} < 1.5$~GeV/$c$.  The authors~\cite{Shen:2016zpp} 
conclude that hadronic rescattering plays a modest but important role in 
the system development and particle species dependence of $v_{2}$ in 
these small systems.

Figure~\ref{fig:figure_wampt} compares the experimental data to 
transport model calculations of $v_{2}(p_{T})$ for each system using 
\textsc{ampt}~\cite{lin_multiphase_2005}. The \textsc{ampt} model has 
been successful in describing various features of small-system 
collectivity at RHIC and the LHC, over a wide range of collision 
energies~\cite{Adare:2015cpn,Koop:2015wea,Koop:2015trj,Ma:2016fve,ma_long-range_2014,ma_long-range_2014}. 
It uses Monte Carlo Glauber initial conditions, and it models the 
evolution of the system via strings that melt into partons, followed by 
a succession of partonic scattering, spatial coalescence, and late-stage 
hadronic scattering implemented in \textsc{art}~\cite{PhysRevC.52.2037}.  
We show results from the full \textsc{ampt} time evolution with a 
partonic cross section $\sigma_{\rm part} = 1.5$ mb, as well as results with 
the hadronic rescattering turned off.  We calculate $v_2$ in central 
(i.e., $b < 2$ fm) \textsc{ampt} events, relative to the parton 
participant plane. That is, the event plane is calculated using the 
initial coordinates of the partons, as they emerge from string melting 
at early times. We observe that the full \textsc{ampt} describes the 
mass-dependent splitting in \dau and \hau for $p_{T} < 1.5$~GeV/$c$. In 
\pau collisions, the model results in a smaller mass splitting, which is 
reversed at high $p_T$ yet below the experimental data. As noted in 
~\cite{Li:2016flp}, \textsc{ampt} generates significant $v_{2}$, and in 
particular mass splitting, in the hadronic rescattering stage. As also 
shown in Figure~\ref{fig:figure_wampt}, the results without rescattering 
have significantly lower $v_{2}$ values and almost no mass splitting for 
$p_{T}<1$~GeV/$c$.  At higher $p_{T}$, the feature of $v_{2}$ for 
protons being greater than pions remains without hadronic rescattering 
and is associated with the spatial coalescence implementation for 
hadronization.

\begin{figure*}[tbh]
  \includegraphics[width=1.0\linewidth]{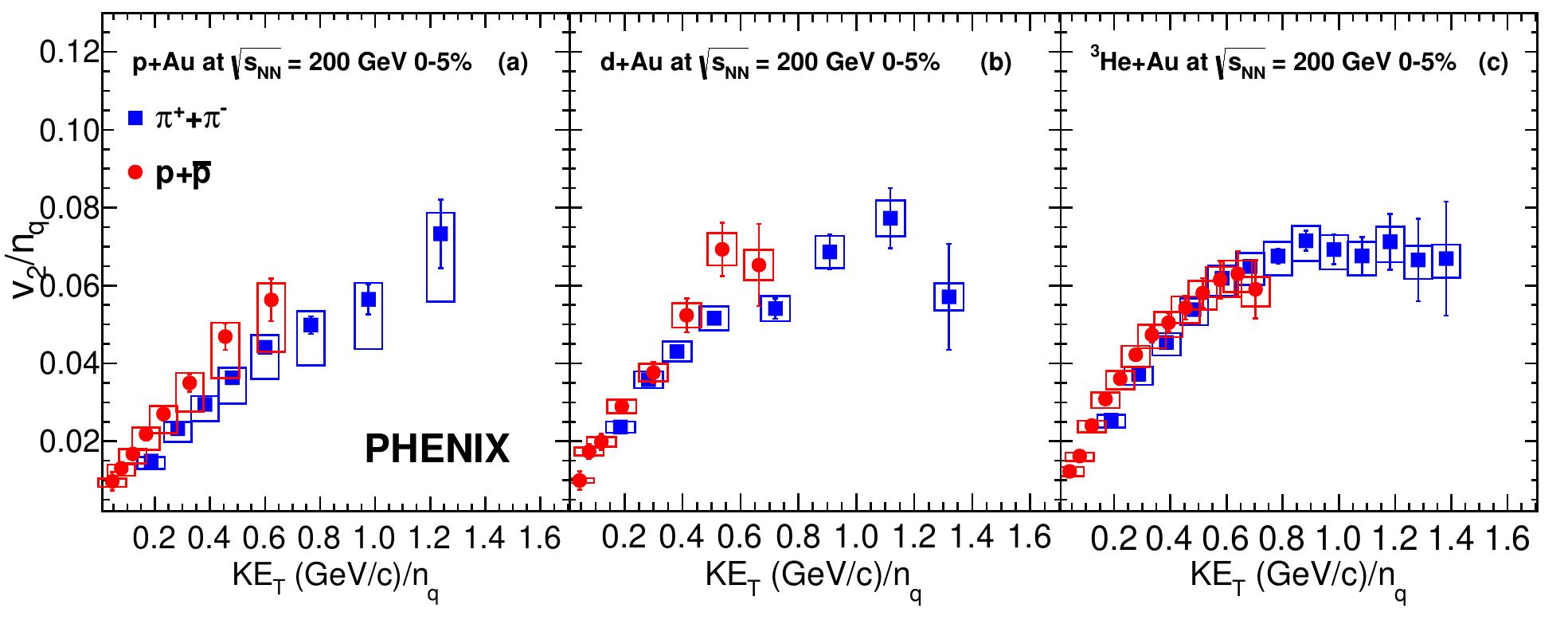}
  \caption{Scaling of $v_2(p_T)$ with the number of constituent quarks 
in each hadron species, in 0\%-5\% central (a) \pau, (b) \dau, and (c) 
\hau collisions at \sqsn = 200 GeV.}
\label{fig:figure_quarkscaling}
\end{figure*}

Figure~\ref{fig:figure_ratio} shows the ratio of pion to proton 
$v_2(p_T)$ in all collision systems, with the corresponding theory 
curves overlaid. In the ratio, many systematic uncertainties cancel and 
thus one sees more precisely that the data exhibit a similar trend in 
all collisions systems where pion $v_{2}$ is larger than proton $v_{2}$ 
for $p_{T} < 1.5$~GeV/$c$, with the order reversed at higher $p_T$. 
Linear fits on these ratios ranged from 0.5 GeV/$c$ to 3.0~GeV/$c$ , 
which include both the statistical and the systematic uncertainties, 
yield slope values of $-0.22\pm 0.07$ in \pau 
collisions,$-0.40~\pm~0.07$ in \dau collisions, and $-0.34~\pm~0.03$ in 
\hau collisions.  In this ratio, one can clearly see that \supersonic, 
i\textsc{ebe-vishnu}, and the full \textsc{ampt} modeling describe the 
mass splitting in \dau and \hau for $p_{T} < 1.5$~GeV/$c$.  In the \pau 
case, it appears that the calculations over-predict the more modest 
splitting at the lowest measured $p_{T} = 0.5$~GeV/$c$.  The results 
from \textsc{ampt} without hadronic rescattering have very little mass 
splitting at low $p_{T}$ in disagreement with the experimental data, 
particularly for \dau and \hau collisions. Above the crossing point, 
\supersonic, and i\textsc{ebe-vishnu} predict nearly flat ratios, while 
\textsc{ampt} describes the ratio of the $v_2$ values, but not their 
individual magnitudes. These differences may be attributed to the 
different hadronization mechanisms (e.g. - if recombination is included) 
in the models.

The observation of a mass-dependent $v_{2}$ strengthens the case for 
associating small-system collectivity with the expansion of QGP droplets 
formed in these collisions, where the splitting can be understood in 
terms of the presence of a common radial flow field with anisotropic 
modulations driven by initial geometry. However, the theoretical 
calculations presented in this paper provide several alternative 
explanations of how the azimuthal anisotropies for different particle 
species may occur. For instance, in kinetic transport, parton scattering 
translates initial geometry into final state momentum anisotropy, but it 
does not account for the observed mass splitting. Instead, this feature 
has been shown to arise solely from the hadronic rescattering stage 
where different hadrons have different inelastic cross 
sections~\cite{Li:2016flp}. There is more hadronic rescattering in \hau 
and \dau compared with \pau for these central collisions because they have 
a higher particle density.  It is interesting that this conclusion based 
on \textsc{ampt} regarding the contribution of the hadronic rescattering 
stage is opposite to that reached using viscous 
hydrodynamics~\cite{Shen:2016zpp}. Differences in the hadronic 
scattering packages B3D~\cite{Novak:2013bqa} used in \supersonic, 
\textsc{urqmd}~\cite{Bleicher:1999xi} used in i\textsc{ebe-vishnu}, and 
\textsc{art}~\cite{lin_multiphase_2005} used in \textsc{ampt} warrant 
further investigation.

Finally, we return to the high $p_{T}$ region where neither viscous 
hydrodynamics nor parton transport calculations match the data. 
Figure~\ref{fig:figure_quarkscaling} shows the scaling of $v_2$ with 
constituent quarks as a function of transverse kinetic energy per quark 
$KE_{T}/n_{q} = (\sqrt{p_{T}^{2}+m^{2}}-m)/n_{q}$, where $m$ is the mass 
of the hadron and $n_{q}$ represents the number of constituent quarks in 
the hadron. In all three systems, the $v_{2}/n_{q}$ for pions and 
protons as a function of $KE_{T}/n_{q}$ follow an approximate 
quark-number scaling. The same scaling was previously observed in $A$$+$$A$ 
collisions~\cite{Adler:2003kt,Adare:2006ti,PhysRevLett.87.182301,PhysRevLett.92.052302}. At intermediate $p_T$~(1.5--4~GeV/$c$), 
the enhancement of baryons over mesons and the reversed mass ordering of 
$v_{2}$ in $A$$+$$A$ collisions have been interpreted in terms of 
hadronization via recombination. At even higher $p_{T}$, the scaling 
breaks down in noncentral $A$$+$$A$ collisions~\cite{Adare:2012vq}. Similar 
to the observations in $A$$+$$A$, the enhancement of baryon over meson yields 
at intermediate $p_{T}$ has been observed in central \dau 
collisions~\cite{Adare:2013esx}, and now we also see the scaling with 
$n_{q}$ in all three small collision systems. The scaling works better 
in \dau and \hau collisions, where the projectile sizes and the particle 
densities are higher.

\section{Summary}

We have presented results on the transverse momentum dependence of 
elliptic flow $v_2$ of identified pions and (anti)protons in central 
0\%--5\% \pau, \dau and \hau at \sqsn = 200 GeV.  The \dau and \hau data 
show a clear mass splitting with $v_{2}$ for pions larger than $v_{2}$ 
of protons for $p_{T}<1.5$~GeV/$c$, and then a reversal of the ordering 
at higher $p_{T}$. The mass dependence is smaller in \pau collisions 
than in \dau and \hau collisions.  Theoretical calculations, from 
viscous hydrodynamics and parton transport, yield a reasonable 
description of the low $p_{T}$ mass splitting, despite having quite 
different mechanisms responsible for the observed mass dependence. At 
higher $p_{T}$, both models fail to describe the data, missing either 
the absolute value or the observed mass dependence.  A scaling of 
$v_{2}$ with the number of constituent quarks, motivated by 
recombination, is observed in the data and is found to hold better in 
\dau and \hau collisions, where the particle multiplicities are larger. 
All of these observations are qualitatively similar to previously 
measured effects in $A$$+$$A$ collisions. This again puts into sharp focus the 
question of whether the observations can be understood as arising from 
the same underlying physics, e.g. inviscid fluid expansion, in both 
large and small collisions systems. While alternative physics mechanisms have been proposed, detailed comparisons with the experimental results are not yet available. This paper provides important constraints on the mass dependence of the  particle correlations in small collision systems.  

\section*{ACKNOWLEDGMENTS}   

We thank the staff of the Collider-Accelerator and Physics
Departments at Brookhaven National Laboratory and the staff of
the other PHENIX participating institutions for their vital
contributions.  We acknowledge support from the 
Office of Nuclear Physics in the
Office of Science of the Department of Energy,
the National Science Foundation, 
Abilene Christian University Research Council, 
Research Foundation of SUNY, and
Dean of the College of Arts and Sciences, Vanderbilt University 
(U.S.A),
Ministry of Education, Culture, Sports, Science, and Technology
and the Japan Society for the Promotion of Science (Japan),
Conselho Nacional de Desenvolvimento Cient\'{\i}fico e
Tecnol{\'o}gico and Funda\c c{\~a}o de Amparo {\`a} Pesquisa do
Estado de S{\~a}o Paulo (Brazil),
Natural Science Foundation of China (People's Republic of China),
Croatian Science Foundation and
Ministry of Science and Education (Croatia),
Ministry of Education, Youth and Sports (Czech Republic),
Centre National de la Recherche Scientifique, Commissariat
{\`a} l'{\'E}nergie Atomique, and Institut National de Physique
Nucl{\'e}aire et de Physique des Particules (France),
Bundesministerium f\"ur Bildung und Forschung, Deutscher
Akademischer Austausch Dienst, and Alexander von Humboldt Stiftung (Germany),
J. Bolyai Research Scholarship, EFOP, the New National Excellence
Program ({\'U}NKP), NKFIH, and OTKA (Hungary),
Department of Atomic Energy and Department of Science and Technology (India), 
Israel Science Foundation (Israel), 
Basic Science Research Program through NRF of the Ministry of Education (Korea),
Physics Department, Lahore University of Management Sciences (Pakistan),
Ministry of Education and Science, Russian Academy of Sciences,
Federal Agency of Atomic Energy (Russia),
VR and Wallenberg Foundation (Sweden), 
the U.S. Civilian Research and Development Foundation for the
Independent States of the Former Soviet Union, 
the Hungarian American Enterprise Scholarship Fund,
the US-Hungarian Fulbright Foundation,
and the US-Israel Binational Science Foundation.

\section*{APPENDIX:  DATA TABLES}

Tables \ref{t:pAupion} and \ref{t:HeAupion} show the values of 
$v_2(\pt)$ for pions, kaons, and protons in central 0\%--5\% \pau and 
\hau collisions at \sqsn = 200 GeV.

\begin{table}[tbh]
\caption{\label{t:pAupion}Values of $v_2(\pt)$ for pions, kaons, and 
protons in central 0\%--5\% \pau collisions at \sqsn = 200 GeV.
}
\begin{ruledtabular}  \begin{tabular}{cccccc}
      &\pt range (GeV/$c$) & $v_2$ & $\pm$ stat & + syst & - syst \\ 
\hline
\multirow{8}{*}{$\pi^{+}+\pi^{-}$} \\ 
& 0.40--0.60 & 0.030 & 0.001 & 0.002 & 0.004 \\
&  0.60--0.80 & 0.046 & 0.002 & 0.003 & 0.007 \\
& 0.80--1.00 & 0.059 & 0.002 & 0.004 & 0.009 \\
& 1.00--1.20 & 0.073 & 0.003 & 0.005 & 0.013 \\
& 1.20--1.50 & 0.088 & 0.003 & 0.006 & 0.016 \\
& 1.50--1.90 & 0.100 & 0.005 & 0.007 & 0.021 \\
& 1.90--2.40 & 0.113 & 0.008 & 0.008 & 0.025 \\
& 2.40--3.00 & 0.147 & 0.018 & 0.011 & 0.035 \\
\multirow{6}{*}{$K^{+}+K^{-}$} \\
& 0.40--0.60 & 0.022 & 0.006 & 0.002 & 0.003 \\ 
& 0.60--0.80 & 0.037 & 0.005 & 0.003 & 0.005 \\
& 0.80--1.00 & 0.056 & 0.006 & 0.004 & 0.008 \\
& 1.00--1.20 & 0.068 & 0.007 & 0.005 & 0.012 \\
& 1.20--1.50 & 0.079 & 0.007 & 0.006 & 0.015 \\
& 1.50--1.90 & 0.091 & 0.009 & 0.007 & 0.019 \\
\multirow{8}{*}{$p+\bar{p}$} \\
& 0.40--0.60 & 0.029 & 0.007 & 0.002 & 0.004 \\
& 0.60--0.80 & 0.039 & 0.005 & 0.003 & 0.006 \\
& 0.80--1.00 & 0.050 & 0.005 & 0.004 & 0.007 \\
& 1.00--1.20 & 0.066 & 0.005 & 0.005 & 0.012 \\
& 1.20--1.50 & 0.081 & 0.005 & 0.006 & 0.015 \\
& 1.50--1.90 & 0.105 & 0.007 & 0.008 & 0.022 \\
& 1.90--2.40 & 0.141 & 0.010 & 0.011 & 0.032 \\
& 2.40--3.00 & 0.169 & 0.016 & 0.013 & 0.040 \\
\end{tabular} \end{ruledtabular}
\end{table}

\begin{table}[htb]
\caption{\label{t:HeAupion}Values of $v_2(\pt)$ for pions, kaons, and 
protons in central 0\%--5\% \hau collisions at \sqsn = 200 GeV.}
\begin{ruledtabular} \begin{tabular}{cccccc}
     &\pt range (GeV/$c$) & $v_2$ & $\pm$ stat & + syst & - syst \\
\hline
\multirow{13}{*}{$\pi^{+}+\pi^{-}$} \\
& 0.40--0.60 & 0.051 & 0.001 & 0.003 & 0.004 \\
& 0.60--0.80 & 0.074 & 0.001 & 0.004 & 0.005 \\
& 0.80--1.00 & 0.091 & 0.001 & 0.005 & 0.007 \\
& 1.00--1.20 & 0.108 & 0.002 & 0.006 & 0.008 \\
& 1.20--1.40 & 0.124 & 0.002 & 0.007 & 0.009 \\
& 1.40--1.60 & 0.130 & 0.003 & 0.007 & 0.010 \\
& 1.60--1.80 & 0.135 & 0.004 & 0.007 & 0.010 \\
& 1.80--2.00 & 0.143 & 0.005 & 0.008 & 0.011 \\
& 2.00--2.20 & 0.138 & 0.008 & 0.007 & 0.010 \\
& 2.20--2.40 & 0.135 & 0.010 & 0.007 & 0.010 \\
& 2.40--2.60 & 0.142 & 0.014 & 0.008 & 0.010 \\
& 2.60--2.80 & 0.133 & 0.021 & 0.007 & 0.010 \\
& 2.80--3.00 & 0.134 & 0.029 & 0.007 & 0.010 \\
\multirow{8}{*}{$K^{+}+K^{-}$} \\
& 0.40--0.60 & 0.041 & 0.003 & 0.002 & 0.003 \\
& 0.60--0.80 & 0.054 & 0.003 & 0.003 & 0.004 \\
& 0.80--1.00 & 0.077 & 0.003 & 0.004 & 0.006 \\
& 1.00--1.20 & 0.093 & 0.004 & 0.005 & 0.007 \\
& 1.20--1.40 & 0.109 & 0.005 & 0.006 & 0.008 \\
& 1.40--1.60 & 0.115 & 0.006 & 0.006 & 0.008 \\
& 1.60--1.80 & 0.123 & 0.007 & 0.007 & 0.009 \\
& 1.80--2.00 & 0.142 & 0.009 & 0.008 & 0.010 \\
\multirow{13}{*}{$p+\bar{p}$} \\
& 0.40--0.60 & 0.037 & 0.002 & 0.003 & 0.004 \\
& 0.60--0.80 & 0.049 & 0.002 & 0.003 & 0.004 \\
& 0.80--1.00 & 0.072 & 0.002 & 0.004 & 0.005 \\
& 1.00--1.20 & 0.093 & 0.003 & 0.005 & 0.007 \\
& 1.20--1.40 & 0.108 & 0.004 & 0.006 & 0.008 \\
& 1.40--1.60 & 0.127 & 0.005 & 0.007 & 0.009 \\
& 1.60--1.80 & 0.142 & 0.006 & 0.008 & 0.010 \\
& 1.80--2.00 & 0.151 & 0.007 & 0.008 & 0.011 \\
& 2.00--2.20 & 0.163 & 0.009 & 0.009 & 0.012 \\
& 2.20--2.40 & 0.174 & 0.012 & 0.009 & 0.013 \\
& 2.40--2.60 & 0.184 & 0.014 & 0.010 & 0.014 \\
& 2.60--2.80 & 0.189 & 0.018 & 0.010 & 0.014 \\
& 2.80--3.00 & 0.177 & 0.023 & 0.010 & 0.013 \\    
    \end{tabular} \end{ruledtabular}
 \end{table}

\clearpage


\begin{thebibliography}{59}%
\makeatletter
\providecommand \@ifxundefined [1]{%
 \@ifx{#1\undefined}
}%
\providecommand \@ifnum [1]{%
 \ifnum #1\expandafter \@firstoftwo
 \else \expandafter \@secondoftwo
 \fi
}%
\providecommand \@ifx [1]{%
 \ifx #1\expandafter \@firstoftwo
 \else \expandafter \@secondoftwo
 \fi
}%
\providecommand \natexlab [1]{#1}%
\providecommand \enquote  [1]{``#1''}%
\providecommand \bibnamefont  [1]{#1}%
\providecommand \bibfnamefont [1]{#1}%
\providecommand \citenamefont [1]{#1}%
\providecommand \href@noop [0]{\@secondoftwo}%
\providecommand \href [0]{\begingroup \@sanitize@url \@href}%
\providecommand \@href[1]{\@@startlink{#1}\@@href}%
\providecommand \@@href[1]{\endgroup#1\@@endlink}%
\providecommand \@sanitize@url [0]{\catcode `\\12\catcode `\$12\catcode
  `\&12\catcode `\#12\catcode `\^12\catcode `\_12\catcode `\%12\relax}%
\providecommand \@@startlink[1]{}%
\providecommand \@@endlink[0]{}%
\providecommand \url  [0]{\begingroup\@sanitize@url \@url }%
\providecommand \@url [1]{\endgroup\@href {#1}{\urlprefix }}%
\providecommand \urlprefix  [0]{URL }%
\providecommand \Eprint [0]{\href }%
\providecommand \doibase [0]{http://dx.doi.org/}%
\providecommand \selectlanguage [0]{\@gobble}%
\providecommand \bibinfo  [0]{\@secondoftwo}%
\providecommand \bibfield  [0]{\@secondoftwo}%
\providecommand \translation [1]{[#1]}%
\providecommand \BibitemOpen [0]{}%
\providecommand \bibitemStop [0]{}%
\providecommand \bibitemNoStop [0]{.\EOS\space}%
\providecommand \EOS [0]{\spacefactor3000\relax}%
\providecommand \BibitemShut  [1]{\csname bibitem#1\endcsname}%
\let\auto@bib@innerbib\@empty
\bibitem [{\citenamefont {Heinz}\ and\ \citenamefont
  {Snellings}(2013)}]{Heinz:2013th}%
  \BibitemOpen
  \bibfield  {author} {\bibinfo {author} {\bibfnamefont {U.}~\bibnamefont
  {Heinz}}\ and\ \bibinfo {author} {\bibfnamefont {R.}~\bibnamefont
  {Snellings}},\ }\bibfield  {title} {\enquote {\bibinfo {title} {{Collective
  flow and viscosity in relativistic heavy-ion collisions}},}\ }\href {\doibase
  10.1146/annurev-nucl-102212-170540} {\bibfield  {journal} {\bibinfo
  {journal} {Ann. Rev. Nucl. Part. Sci.}\ }\textbf {\bibinfo {volume} {63}},\
  \bibinfo {pages} {123} (\bibinfo {year} {2013})}\BibitemShut {NoStop}%
\bibitem [{\citenamefont {Adare}\ \emph
  {et~al.}(2013{\natexlab{a}})\citenamefont {Adare} \emph
  {et~al.}}]{PhysRevLett.111.212301}%
  \BibitemOpen
  \bibfield  {author} {\bibinfo {author} {\bibfnamefont {A.}~\bibnamefont
  {Adare}} \emph {et~al.} (\bibinfo {collaboration} {{PHENIX Collaboration}}),\
  }\bibfield  {title} {\enquote {\bibinfo {title} {{Quadrupole Anisotropy in
  Dihadron Azimuthal Correlations in Central $d$+Au Collisions at
  $\sqrt{s_{NN}}=200$ GeV}},}\ }\href {\doibase 10.1103/PhysRevLett.111.212301}
  {\bibfield  {journal} {\bibinfo  {journal} {Phys. Rev. Lett.}\ }\textbf
  {\bibinfo {volume} {111}},\ \bibinfo {pages} {212301} (\bibinfo {year}
  {2013}{\natexlab{a}})}\BibitemShut {NoStop}%
\bibitem [{\citenamefont {Adare}\ \emph
  {et~al.}(2015{\natexlab{a}})\citenamefont {Adare} \emph
  {et~al.}}]{Adare:2014keg}%
  \BibitemOpen
  \bibfield  {author} {\bibinfo {author} {\bibfnamefont {A.}~\bibnamefont
  {Adare}} \emph {et~al.} (\bibinfo {collaboration} {{PHENIX Collaboration}}),\
  }\bibfield  {title} {\enquote {\bibinfo {title} {{Measurement of long-range
  angular correlation and quadrupole anisotropy of pions and (anti)protons in
  central $d$$+$Au collisions at $\sqrt{s_{_{NN}}}$=200~GeV}},}\ }\href
  {\doibase 10.1103/PhysRevLett.114.192301} {\bibfield  {journal} {\bibinfo
  {journal} {Phys. Rev. Lett.}\ }\textbf {\bibinfo {volume} {114}},\ \bibinfo
  {pages} {192301} (\bibinfo {year} {2015}{\natexlab{a}})}\BibitemShut
  {NoStop}%
\bibitem [{\citenamefont {Adare}\ \emph
  {et~al.}(2015{\natexlab{b}})\citenamefont {Adare} \emph
  {et~al.}}]{Adare:2015ctn}%
  \BibitemOpen
  \bibfield  {author} {\bibinfo {author} {\bibfnamefont {A.}~\bibnamefont
  {Adare}} \emph {et~al.} (\bibinfo {collaboration} {{PHENIX Collaboration}}),\
  }\bibfield  {title} {\enquote {\bibinfo {title} {{Measurements of elliptic
  and triangular flow in high-multiplicity $^{3}$He$+$Au collisions at
  $\sqrt{s_{_{NN}}}=200$ GeV}},}\ }\href {\doibase
  10.1103/PhysRevLett.115.142301} {\bibfield  {journal} {\bibinfo  {journal}
  {Phys. Rev. Lett.}\ }\textbf {\bibinfo {volume} {115}},\ \bibinfo {pages}
  {142301} (\bibinfo {year} {2015}{\natexlab{b}})}\BibitemShut {NoStop}%
\bibitem [{\citenamefont {Abelev}\ \emph
  {et~al.}(2013{\natexlab{a}})\citenamefont {Abelev} \emph
  {et~al.}}]{alice_long_2013}%
  \BibitemOpen
  \bibfield  {author} {\bibinfo {author} {\bibfnamefont {B.}~\bibnamefont
  {Abelev}} \emph {et~al.} (\bibinfo {collaboration} {{ALICE Collaboration}}),\
  }\bibfield  {title} {\enquote {\bibinfo {title} {{Long-range angular
  correlations on the near and away side in $p$Pb collisions at}},}\ }\href
  {\doibase http://dx.doi.org/10.1016/j.physletb.2013.01.012} {\bibfield
  {journal} {\bibinfo  {journal} {Phys. Lett. B}\ }\textbf {\bibinfo {volume}
  {719}},\ \bibinfo {pages} {29} (\bibinfo {year}
  {2013}{\natexlab{a}})}\BibitemShut {NoStop}%
\bibitem [{\citenamefont {Aad}\ \emph {et~al.}(2013)\citenamefont {Aad} \emph
  {et~al.}}]{atlas_observation_2012}%
  \BibitemOpen
  \bibfield  {author} {\bibinfo {author} {\bibfnamefont {G.}~\bibnamefont
  {Aad}} \emph {et~al.} (\bibinfo {collaboration} {{ATLAS Collaboration}}),\
  }\bibfield  {title} {\enquote {\bibinfo {title} {{Oservation of Associated
  Near-Side and Away-Side Long-Range Correlations in $\sqrt{s_{NN}}=5.02$ TeV
  Proton-Lead Collisions with the ATLAS Detector}},}\ }\href {\doibase
  10.1103/PhysRevLett.110.182302} {\bibfield  {journal} {\bibinfo  {journal}
  {Phys. Rev. Lett.}\ }\textbf {\bibinfo {volume} {110}},\ \bibinfo {pages}
  {182302} (\bibinfo {year} {2013})}\BibitemShut {NoStop}%
\bibitem [{\citenamefont {Chatrchyan}\ \emph {et~al.}(2013)\citenamefont
  {Chatrchyan} \emph {et~al.}}]{cms_observation_2012}%
  \BibitemOpen
  \bibfield  {author} {\bibinfo {author} {\bibfnamefont {S.}~\bibnamefont
  {Chatrchyan}} \emph {et~al.} (\bibinfo {collaboration} {{CMS
  Collaboration}}),\ }\bibfield  {title} {\enquote {\bibinfo {title}
  {{Observation of long-range, near-side angular correlations in $p$Pb
  collisions at the LHC}},}\ }\href {\doibase
  http://dx.doi.org/10.1016/j.physletb.2012.11.025} {\bibfield  {journal}
  {\bibinfo  {journal} {Phys. Lett. B}\ }\textbf {\bibinfo {volume} {718}},\
  \bibinfo {pages} {795} (\bibinfo {year} {2013})}\BibitemShut {NoStop}%
\bibitem [{\citenamefont {Khachatryan}\ \emph {et~al.}(2016)\citenamefont
  {Khachatryan} \emph {et~al.}}]{Khachatryan:2015lva}%
  \BibitemOpen
  \bibfield  {author} {\bibinfo {author} {\bibfnamefont {V.}~\bibnamefont
  {Khachatryan}} \emph {et~al.} (\bibinfo {collaboration} {{CMS
  Collaboration}}),\ }\bibfield  {title} {\enquote {\bibinfo {title}
  {{Measurement of long-range near-side two-particle angular correlations in
  $pp$ collisions at $\sqrt{s}=$13 TeV}},}\ }\href {\doibase
  10.1103/PhysRevLett.116.172302} {\bibfield  {journal} {\bibinfo  {journal}
  {Phys. Rev. Lett.}\ }\textbf {\bibinfo {volume} {116}},\ \bibinfo {pages}
  {172302} (\bibinfo {year} {2016})}\BibitemShut {NoStop}%
\bibitem [{\citenamefont {Aad}\ \emph {et~al.}(2016)\citenamefont {Aad} \emph
  {et~al.}}]{Aad:2015gqa}%
  \BibitemOpen
  \bibfield  {author} {\bibinfo {author} {\bibfnamefont {G.}~\bibnamefont
  {Aad}} \emph {et~al.} (\bibinfo {collaboration} {{ATLAS Collaboration}}),\
  }\bibfield  {title} {\enquote {\bibinfo {title} {{Observation of Long-Range
  Elliptic Azimuthal Anisotropies in $\sqrt{s}=$13 and 2.76 TeV $pp$ Collisions
  with the ATLAS Detector}},}\ }\href {\doibase 10.1103/PhysRevLett.116.172301}
  {\bibfield  {journal} {\bibinfo  {journal} {Phys. Rev. Lett.}\ }\textbf
  {\bibinfo {volume} {116}},\ \bibinfo {pages} {172301} (\bibinfo {year}
  {2016})}\BibitemShut {NoStop}%
\bibitem [{\citenamefont {Khachatryan}\ \emph {et~al.}()\citenamefont
  {Khachatryan} \emph {et~al.}}]{Khachatryan:2010gv}%
  \BibitemOpen
  \bibfield  {author} {\bibinfo {author} {\bibfnamefont {Vardan}\ \bibnamefont
  {Khachatryan}} \emph {et~al.} (\bibinfo {collaboration} {{CMS
  Collaboration}}),\ }\href@noop {} {\enquote {\bibinfo {title} {{Observation
  of Long-Range Near-Side Angular Correlations in Proton-Proton Collisions at
  the LHC}},}\ }\bibinfo {note} {{J. High Energy Phys. {\bf 09 (2010)}
  091}}\BibitemShut {NoStop}%
\bibitem [{\citenamefont {Khachatryan}\ \emph {et~al.}(2017)\citenamefont
  {Khachatryan} \emph {et~al.}}]{Khachatryan:2016txc}%
  \BibitemOpen
  \bibfield  {author} {\bibinfo {author} {\bibfnamefont {V.}~\bibnamefont
  {Khachatryan}} \emph {et~al.} (\bibinfo {collaboration} {{CMS
  Collaboration}}),\ }\bibfield  {title} {\enquote {\bibinfo {title} {{Evidence
  for collectivity in $pp$ collisions at the LHC}},}\ }\href {\doibase
  10.1016/j.physletb.2016.12.009} {\bibfield  {journal} {\bibinfo  {journal}
  {Phys. Lett. B}\ }\textbf {\bibinfo {volume} {765}},\ \bibinfo {pages} {193}
  (\bibinfo {year} {2017})}\BibitemShut {NoStop}%
\bibitem [{\citenamefont {Dusling}\ and\ \citenamefont
  {Venugopalan}(2012)}]{Dusling:2012iga}%
  \BibitemOpen
  \bibfield  {author} {\bibinfo {author} {\bibfnamefont {K.}~\bibnamefont
  {Dusling}}\ and\ \bibinfo {author} {\bibfnamefont {R.}~\bibnamefont
  {Venugopalan}},\ }\bibfield  {title} {\enquote {\bibinfo {title} {{Azimuthal
  collimation of long range rapidity correlations by strong color fields in
  high multiplicity hadron-hadron collisions}},}\ }\href {\doibase
  10.1103/PhysRevLett.108.262001} {\bibfield  {journal} {\bibinfo  {journal}
  {Phys. Rev. Lett.}\ }\textbf {\bibinfo {volume} {108}},\ \bibinfo {pages}
  {262001} (\bibinfo {year} {2012})}\BibitemShut {NoStop}%
\bibitem [{\citenamefont {Ortiz~Velasquez}\ \emph {et~al.}(2013)\citenamefont
  {Ortiz~Velasquez}, \citenamefont {Christiansen}, \citenamefont
  {Cuautle~Flores}, \citenamefont {Maldonado~Cervantes},\ and\ \citenamefont
  {Paic}}]{Ortiz:2013yxa}%
  \BibitemOpen
  \bibfield  {author} {\bibinfo {author} {\bibfnamefont {A.}~\bibnamefont
  {Ortiz~Velasquez}}, \bibinfo {author} {\bibfnamefont {P.}~\bibnamefont
  {Christiansen}}, \bibinfo {author} {\bibfnamefont {E.}~\bibnamefont
  {Cuautle~Flores}}, \bibinfo {author} {\bibfnamefont {I.~A.}\ \bibnamefont
  {Maldonado~Cervantes}}, \ and\ \bibinfo {author} {\bibfnamefont
  {G.}~\bibnamefont {Paic}},\ }\bibfield  {title} {\enquote {\bibinfo {title}
  {{Color Reconnection and Flowlike Patterns in $pp$ Collisions}},}\ }\href
  {\doibase 10.1103/PhysRevLett.111.042001} {\bibfield  {journal} {\bibinfo
  {journal} {Phys. Rev. Lett.}\ }\textbf {\bibinfo {volume} {111}},\ \bibinfo
  {pages} {042001} (\bibinfo {year} {2013})}\BibitemShut {NoStop}%
\bibitem [{\citenamefont {Aidala}\ \emph {et~al.}(2017)\citenamefont {Aidala}
  \emph {et~al.}}]{PhysRevC.95.034910}%
  \BibitemOpen
  \bibfield  {author} {\bibinfo {author} {\bibfnamefont {C.}~\bibnamefont
  {Aidala}} \emph {et~al.} (\bibinfo {collaboration} {PHENIX Collaboration}),\
  }\bibfield  {title} {\enquote {\bibinfo {title} {{Measurement of long-range
  angular correlations and azimuthal anisotropies in high-multiplicity $p$$+$Au
  collisions at $\sqrt{{s}_{\mathit{NN}}}=200$ GeV}},}\ }\href {\doibase
  10.1103/PhysRevC.95.034910} {\bibfield  {journal} {\bibinfo  {journal} {Phys.
  Rev. C}\ }\textbf {\bibinfo {volume} {95}},\ \bibinfo {pages} {034910}
  (\bibinfo {year} {2017})}\BibitemShut {NoStop}%
\bibitem [{\citenamefont {Nagle}\ \emph {et~al.}(2014)\citenamefont {Nagle},
  \citenamefont {Adare}, \citenamefont {Beckman}, \citenamefont {Koblesky},
  \citenamefont {Koop}, \citenamefont {McGlinchey}, \citenamefont {Romatschke},
  \citenamefont {Carlson}, \citenamefont {Lynn},\ and\ \citenamefont
  {McCumber}}]{nagle_exploiting_2013}%
  \BibitemOpen
  \bibfield  {author} {\bibinfo {author} {\bibfnamefont {J.~L.}\ \bibnamefont
  {Nagle}}, \bibinfo {author} {\bibfnamefont {A.}~\bibnamefont {Adare}},
  \bibinfo {author} {\bibfnamefont {S.}~\bibnamefont {Beckman}}, \bibinfo
  {author} {\bibfnamefont {T.}~\bibnamefont {Koblesky}}, \bibinfo {author}
  {\bibfnamefont {J.~O.}\ \bibnamefont {Koop}}, \bibinfo {author}
  {\bibfnamefont {D.}~\bibnamefont {McGlinchey}}, \bibinfo {author}
  {\bibfnamefont {P.}~\bibnamefont {Romatschke}}, \bibinfo {author}
  {\bibfnamefont {J.}~\bibnamefont {Carlson}}, \bibinfo {author} {\bibfnamefont
  {J.~E.}\ \bibnamefont {Lynn}}, \ and\ \bibinfo {author} {\bibfnamefont
  {M.}~\bibnamefont {McCumber}},\ }\bibfield  {title} {\enquote {\bibinfo
  {title} {{Exploiting Intrinsic Triangular Geometry in Relativistic
  $^3$He$+$Au Collisions to Disentangle Medium Properties}},}\ }\href
  {http://arxiv.org/abs/1312.4565} {\bibfield  {journal} {\bibinfo  {journal}
  {Phys. Rev. Lett.}\ }\textbf {\bibinfo {volume} {113}},\ \bibinfo {pages}
  {112301} (\bibinfo {year} {2014})}\BibitemShut {NoStop}%
\bibitem [{\citenamefont {Bozek}\ and\ \citenamefont
  {Broniowski}(2015)}]{Bozek:2015qpa}%
  \BibitemOpen
  \bibfield  {author} {\bibinfo {author} {\bibfnamefont {P.}~\bibnamefont
  {Bozek}}\ and\ \bibinfo {author} {\bibfnamefont {W.}~\bibnamefont
  {Broniowski}},\ }\bibfield  {title} {\enquote {\bibinfo {title}
  {{Hydrodynamic modeling of $^3$He$+$Au collisions at $\sqrt {s_{NN}}$=200
  GeV}},}\ }\href {\doibase 10.1016/j.physletb.2015.05.068} {\bibfield
  {journal} {\bibinfo  {journal} {Phys. Lett. B}\ }\textbf {\bibinfo {volume}
  {747}},\ \bibinfo {pages} {135} (\bibinfo {year} {2015})}\BibitemShut
  {NoStop}%
\bibitem [{\citenamefont {Schenke}\ and\ \citenamefont
  {Venugopalan}(2014)}]{Schenke:2014gaa}%
  \BibitemOpen
  \bibfield  {author} {\bibinfo {author} {\bibfnamefont {B.}~\bibnamefont
  {Schenke}}\ and\ \bibinfo {author} {\bibfnamefont {R.}~\bibnamefont
  {Venugopalan}},\ }\bibfield  {title} {\enquote {\bibinfo {title} {{Collective
  effects in light–heavy ion collisions}},}\ }\bibfield  {booktitle} {\emph
  {\bibinfo {booktitle} {{Proceedings, 24th International Conference on
  Ultra-Relativistic Nucleus-Nucleus Collisions (Quark Matter 2014)}}},\ }\href
  {\doibase 10.1016/j.nuclphysa.2014.08.092} {\bibfield  {journal} {\bibinfo
  {journal} {Nucl. Phys. A}\ }\textbf {\bibinfo {volume} {931}},\ \bibinfo
  {pages} {1039} (\bibinfo {year} {2014})}\BibitemShut {NoStop}%
\bibitem [{\citenamefont {Shen}\ \emph {et~al.}(2017)\citenamefont {Shen},
  \citenamefont {Paquet}, \citenamefont {Denicol}, \citenamefont {Jeon},\ and\
  \citenamefont {Gale}}]{Shen:2016zpp}%
  \BibitemOpen
  \bibfield  {author} {\bibinfo {author} {\bibfnamefont {C.}~\bibnamefont
  {Shen}}, \bibinfo {author} {\bibfnamefont {J.~F.}\ \bibnamefont {Paquet}},
  \bibinfo {author} {\bibfnamefont {G.~S.}\ \bibnamefont {Denicol}}, \bibinfo
  {author} {\bibfnamefont {S.}~\bibnamefont {Jeon}}, \ and\ \bibinfo {author}
  {\bibfnamefont {C.}~\bibnamefont {Gale}},\ }\bibfield  {title} {\enquote
  {\bibinfo {title} {{Collectivity and electromagnetic radiation in small
  systems}},}\ }\href {\doibase 10.1103/PhysRevC.95.014906} {\bibfield
  {journal} {\bibinfo  {journal} {Phys. Rev. C}\ }\textbf {\bibinfo {volume}
  {95}},\ \bibinfo {pages} {014906} (\bibinfo {year} {2017})}\BibitemShut
  {NoStop}%
\bibitem [{\citenamefont {Weller}\ and\ \citenamefont
  {Romatschke}(2017)}]{Weller:2017tsr}%
  \BibitemOpen
  \bibfield  {author} {\bibinfo {author} {\bibfnamefont {R.~D.}\ \bibnamefont
  {Weller}}\ and\ \bibinfo {author} {\bibfnamefont {P.}~\bibnamefont
  {Romatschke}},\ }\bibfield  {title} {\enquote {\bibinfo {title} {{One fluid
  to rule them all: viscous hydrodynamic description of event-by-event central
  $p$$+$$p$, $p$$+$Pb and Pb$+$Pb collisions at $\sqrt{s}=5.02$ TeV}},}\ }\href
  {\doibase 10.1016/j.physletb.2017.09.077} {\bibfield  {journal} {\bibinfo
  {journal} {Phys. Lett. B}\ }\textbf {\bibinfo {volume} {774}},\ \bibinfo
  {pages} {351} (\bibinfo {year} {2017})}\BibitemShut {NoStop}%
\bibitem [{\citenamefont {Adler}\ \emph
  {et~al.}(2003{\natexlab{a}})\citenamefont {Adler} \emph
  {et~al.}}]{Adler:2003kt}%
  \BibitemOpen
  \bibfield  {author} {\bibinfo {author} {\bibfnamefont {S.~S.}\ \bibnamefont
  {Adler}} \emph {et~al.} (\bibinfo {collaboration} {{PHENIX Collaboration}}),\
  }\bibfield  {title} {\enquote {\bibinfo {title} {{Elliptic flow of identified
  hadrons in Au$+$Au collisions at $\sqrt{s_{NN}}=200$ GeV}},}\ }\href
  {\doibase 10.1103/PhysRevLett.91.182301} {\bibfield  {journal} {\bibinfo
  {journal} {Phys. Rev. Lett.}\ }\textbf {\bibinfo {volume} {91}},\ \bibinfo
  {pages} {182301} (\bibinfo {year} {2003}{\natexlab{a}})}\BibitemShut
  {NoStop}%
\bibitem [{\citenamefont {Abelev}\ \emph
  {et~al.}(2013{\natexlab{b}})\citenamefont {Abelev} \emph
  {et~al.}}]{ABELEV:2013wsa}%
  \BibitemOpen
  \bibfield  {author} {\bibinfo {author} {\bibfnamefont {B.~B.}\ \bibnamefont
  {Abelev}} \emph {et~al.} (\bibinfo {collaboration} {{ALICE Collaboration}}),\
  }\bibfield  {title} {\enquote {\bibinfo {title} {{Long-range angular
  correlations of $\rm \pi$, K and p in p-Pb collisions at $\sqrt{s_{\rm NN}}$
  = 5.02 TeV}},}\ }\href {\doibase 10.1016/j.physletb.2013.08.024} {\bibfield
  {journal} {\bibinfo  {journal} {Phys. Lett. B}\ }\textbf {\bibinfo {volume}
  {726}},\ \bibinfo {pages} {164} (\bibinfo {year}
  {2013}{\natexlab{b}})}\BibitemShut {NoStop}%
\bibitem [{\citenamefont {Khachatryan}\ \emph {et~al.}(2015)\citenamefont
  {Khachatryan} \emph {et~al.}}]{Khachatryan:2014jra}%
  \BibitemOpen
  \bibfield  {author} {\bibinfo {author} {\bibfnamefont {V.}~\bibnamefont
  {Khachatryan}} \emph {et~al.} (\bibinfo {collaboration} {{CMS
  Collaboration}}),\ }\bibfield  {title} {\enquote {\bibinfo {title}
  {{Long-range two-particle correlations of strange hadrons with charged
  particles in $p$Pb and PbPb collisions at LHC energies}},}\ }\href {\doibase
  10.1016/j.physletb.2015.01.034} {\bibfield  {journal} {\bibinfo  {journal}
  {Phys. Lett. B}\ }\textbf {\bibinfo {volume} {742}},\ \bibinfo {pages} {200}
  (\bibinfo {year} {2015})}\BibitemShut {NoStop}%
\bibitem [{\citenamefont {Lin}\ \emph {et~al.}(2005)\citenamefont {Lin},
  \citenamefont {Ko}, \citenamefont {Li}, \citenamefont {Zhang},\ and\
  \citenamefont {Pal}}]{lin_multiphase_2005}%
  \BibitemOpen
  \bibfield  {author} {\bibinfo {author} {\bibfnamefont {Z.~W.}\ \bibnamefont
  {Lin}}, \bibinfo {author} {\bibfnamefont {C.~M.}\ \bibnamefont {Ko}},
  \bibinfo {author} {\bibfnamefont {B.~A.}\ \bibnamefont {Li}}, \bibinfo
  {author} {\bibfnamefont {B.}~\bibnamefont {Zhang}}, \ and\ \bibinfo {author}
  {\bibfnamefont {S.}~\bibnamefont {Pal}},\ }\bibfield  {title} {\enquote
  {\bibinfo {title} {Multiphase transport model for relativistic heavy ion
  collisions},}\ }\href {\doibase 10.1103/PhysRevC.72.064901} {\bibfield
  {journal} {\bibinfo  {journal} {Phys. Rev. C}\ }\textbf {\bibinfo {volume}
  {72}},\ \bibinfo {pages} {064901} (\bibinfo {year} {2005})}\BibitemShut
  {NoStop}%
\bibitem [{\citenamefont {Li}\ \emph {et~al.}(2016)\citenamefont {Li},
  \citenamefont {He}, \citenamefont {Lin}, \citenamefont {Molnar},
  \citenamefont {Wang},\ and\ \citenamefont {Xie}}]{Li:2016flp}%
  \BibitemOpen
  \bibfield  {author} {\bibinfo {author} {\bibfnamefont {H.}~\bibnamefont
  {Li}}, \bibinfo {author} {\bibfnamefont {L.}~\bibnamefont {He}}, \bibinfo
  {author} {\bibfnamefont {Z.~W.}\ \bibnamefont {Lin}}, \bibinfo {author}
  {\bibfnamefont {D.}~\bibnamefont {Molnar}}, \bibinfo {author} {\bibfnamefont
  {F.}~\bibnamefont {Wang}}, \ and\ \bibinfo {author} {\bibfnamefont
  {W.}~\bibnamefont {Xie}},\ }\bibfield  {title} {\enquote {\bibinfo {title}
  {{Origin of the mass splitting of elliptic anisotropy in a multiphase
  transport model}},}\ }\href {\doibase 10.1103/PhysRevC.93.051901} {\bibfield
  {journal} {\bibinfo  {journal} {Phys. Rev. C}\ }\textbf {\bibinfo {volume}
  {93}},\ \bibinfo {pages} {051901} (\bibinfo {year} {2016})}\BibitemShut
  {NoStop}%
\bibitem [{\citenamefont {Schenke}\ \emph {et~al.}(2016)\citenamefont
  {Schenke}, \citenamefont {Schlichting}, \citenamefont {Tribedy},\ and\
  \citenamefont {Venugopalan}}]{Schenke:2016lrs}%
  \BibitemOpen
  \bibfield  {author} {\bibinfo {author} {\bibfnamefont {B.}~\bibnamefont
  {Schenke}}, \bibinfo {author} {\bibfnamefont {S.}~\bibnamefont
  {Schlichting}}, \bibinfo {author} {\bibfnamefont {P.}~\bibnamefont
  {Tribedy}}, \ and\ \bibinfo {author} {\bibfnamefont {R.}~\bibnamefont
  {Venugopalan}},\ }\bibfield  {title} {\enquote {\bibinfo {title} {{Mass
  ordering of spectra from fragmentation of saturated gluon states in high
  multiplicity proton-proton collisions}},}\ }\href {\doibase
  10.1103/PhysRevLett.117.162301} {\bibfield  {journal} {\bibinfo  {journal}
  {Phys. Rev. Lett.}\ }\textbf {\bibinfo {volume} {117}},\ \bibinfo {pages}
  {162301} (\bibinfo {year} {2016})}\BibitemShut {NoStop}%
\bibitem [{\citenamefont {Werner}\ \emph {et~al.}(2014)\citenamefont {Werner},
  \citenamefont {Bleicher}, \citenamefont {Guiot}, \citenamefont {Karpenko},\
  and\ \citenamefont {Pierog}}]{Werner:2013ipa}%
  \BibitemOpen
  \bibfield  {author} {\bibinfo {author} {\bibfnamefont {K.}~\bibnamefont
  {Werner}}, \bibinfo {author} {\bibfnamefont {M.}~\bibnamefont {Bleicher}},
  \bibinfo {author} {\bibfnamefont {B.}~\bibnamefont {Guiot}}, \bibinfo
  {author} {\bibfnamefont {Iu.}\ \bibnamefont {Karpenko}}, \ and\ \bibinfo
  {author} {\bibfnamefont {T.}~\bibnamefont {Pierog}},\ }\bibfield  {title}
  {\enquote {\bibinfo {title} {{Evidence for Flow from Hydrodynamic Simulations
  of $p$-Pb Collisions at 5.02 TeV from $\nu_2$ Mass Splitting}},}\ }\href
  {\doibase 10.1103/PhysRevLett.112.232301} {\bibfield  {journal} {\bibinfo
  {journal} {Phys. Rev. Lett.}\ }\textbf {\bibinfo {volume} {112}},\ \bibinfo
  {pages} {232301} (\bibinfo {year} {2014})}\BibitemShut {NoStop}%
\bibitem [{\citenamefont {Romatschke}(2015)}]{Romatschke2015}%
  \BibitemOpen
  \bibfield  {author} {\bibinfo {author} {\bibfnamefont {P.}~\bibnamefont
  {Romatschke}},\ }\bibfield  {title} {\enquote {\bibinfo {title}
  {Light-heavy-ion collisions: a window into pre-equilibrium qcd dynamics?}}\
  }\href {\doibase 10.1140/epjc/s10052-015-3509-3} {\bibfield  {journal}
  {\bibinfo  {journal} {Eur. Phys. J. C}\ }\textbf {\bibinfo {volume} {75}},\
  \bibinfo {pages} {305} (\bibinfo {year} {2015})}\BibitemShut {NoStop}%
\bibitem [{\citenamefont {Adcox}\ \emph {et~al.}(2003)\citenamefont {Adcox}
  \emph {et~al.}}]{Adcox2003469}%
  \BibitemOpen
  \bibfield  {author} {\bibinfo {author} {\bibfnamefont {K.}~\bibnamefont
  {Adcox}} \emph {et~al.} (\bibinfo {collaboration} {{PHENIX Collaboration}}),\
  }\bibfield  {title} {\enquote {\bibinfo {title} {{PHENIX detector
  overview}},}\ }\href {\doibase 10.1016/S0168-9002(02)01950-2} {\bibfield
  {journal} {\bibinfo  {journal} {Nucl. Instrum. Meth. Phys. Res., Sec. A}\
  }\textbf {\bibinfo {volume} {499}},\ \bibinfo {pages} {469} (\bibinfo {year}
  {2003})}\BibitemShut {NoStop}%
\bibitem [{\citenamefont {Aidala}\ \emph {et~al.}(2014)\citenamefont {Aidala}
  \emph {et~al.}}]{Aidala:2013vna}%
  \BibitemOpen
  \bibfield  {author} {\bibinfo {author} {\bibfnamefont {C.}~\bibnamefont
  {Aidala}} \emph {et~al.},\ }\bibfield  {title} {\enquote {\bibinfo {title}
  {{The PHENIX Forward Silicon Vertex Detector}},}\ }\href {\doibase
  10.1016/j.nima.2014.04.017} {\bibfield  {journal} {\bibinfo  {journal} {Nucl.
  Instrum. Meth. Phys. Res., Sec. A}\ }\textbf {\bibinfo {volume} {755}},\
  \bibinfo {pages} {44} (\bibinfo {year} {2014})}\BibitemShut {NoStop}%
\bibitem [{\citenamefont {Aizawa}\ \emph {et~al.}(2003)\citenamefont {Aizawa}
  \emph {et~al.}}]{AIZAWA2003508}%
  \BibitemOpen
  \bibfield  {author} {\bibinfo {author} {\bibfnamefont {M.}~\bibnamefont
  {Aizawa}} \emph {et~al.},\ }\bibfield  {title} {\enquote {\bibinfo {title}
  {Phenix central arm particle id detectors},}\ }\href@noop {} {\bibfield
  {journal} {\bibinfo  {journal} {Nucl. Instrum. Methods Phys. Res., Sec. A}\
  }\textbf {\bibinfo {volume} {499}},\ \bibinfo {pages} {508} (\bibinfo {year}
  {2003})}\BibitemShut {NoStop}%
\bibitem [{\citenamefont {Adare}\ \emph
  {et~al.}(2013{\natexlab{b}})\citenamefont {Adare} \emph
  {et~al.}}]{Adare:2013esx}%
  \BibitemOpen
  \bibfield  {author} {\bibinfo {author} {\bibfnamefont {A.}~\bibnamefont
  {Adare}} \emph {et~al.} (\bibinfo {collaboration} {{PHENIX Collaboration}}),\
  }\bibfield  {title} {\enquote {\bibinfo {title} {{Spectra and ratios of
  identified particles in Au+Au and $d$+Au collisions at $\sqrt{s_{NN}}=200$
  GeV}},}\ }\href {\doibase 10.1103/PhysRevC.88.024906} {\bibfield  {journal}
  {\bibinfo  {journal} {Phys. Rev. C}\ }\textbf {\bibinfo {volume} {88}},\
  \bibinfo {pages} {024906} (\bibinfo {year} {2013}{\natexlab{b}})}\BibitemShut
  {NoStop}%
\bibitem [{gea()}]{geant}%
  \BibitemOpen
  \href@noop {} {}\bibinfo {note} {{\sc geant}3.2.1 Manual (CERN, Geneva,
  1993); available at
  http://wwwasdoc.web.cern.ch/wwwasdoc/pdfdir/geant.pdf.5}\BibitemShut
  {NoStop}%
\bibitem [{\citenamefont {Adare}\ \emph
  {et~al.}(2016{\natexlab{a}})\citenamefont {Adare} \emph
  {et~al.}}]{Adare:2015hla}%
  \BibitemOpen
  \bibfield  {author} {\bibinfo {author} {\bibfnamefont {A.}~\bibnamefont
  {Adare}} \emph {et~al.} (\bibinfo {collaboration} {{PHENIX Collaboration}}),\
  }\bibfield  {title} {\enquote {\bibinfo {title} {{Single electron yields from
  semileptonic charm and bottom hadron decays in Au$+$Au collisions at
  $\sqrt{s_{NN}}=200$ GeV}},}\ }\href {\doibase 10.1103/PhysRevC.93.034904}
  {\bibfield  {journal} {\bibinfo  {journal} {Phys. Rev. C}\ }\textbf {\bibinfo
  {volume} {93}},\ \bibinfo {pages} {034904} (\bibinfo {year}
  {2016}{\natexlab{a}})}\BibitemShut {NoStop}%
\bibitem [{\citenamefont {Allen}\ \emph {et~al.}(2003)\citenamefont {Allen}
  \emph {et~al.}}]{Allen2003549}%
  \BibitemOpen
  \bibfield  {author} {\bibinfo {author} {\bibfnamefont {M.~Allen}\
  \bibnamefont {Allen}} \emph {et~al.},\ }\bibfield  {title} {\enquote
  {\bibinfo {title} {{PHENIX inner detectors}},}\ }\href {\doibase
  http://dx.doi.org/10.1016/S0168-9002(02)01956-3} {\bibfield  {journal}
  {\bibinfo  {journal} {Nucl. Instrum. Methods Phys. Res., Sec. A}\ }\textbf
  {\bibinfo {volume} {499}},\ \bibinfo {pages} {549} (\bibinfo {year}
  {2003})}\BibitemShut {NoStop}%
\bibitem [{\citenamefont {Adare}\ \emph {et~al.}(2014)\citenamefont {Adare}
  \emph {et~al.}}]{bbc}%
  \BibitemOpen
  \bibfield  {author} {\bibinfo {author} {\bibfnamefont {A.}~\bibnamefont
  {Adare}} \emph {et~al.} (\bibinfo {collaboration} {{PHENIX Collaboration}}),\
  }\bibfield  {title} {\enquote {\bibinfo {title} {{Centrality categorization
  for $R_{p(d)+Au}$ in high-energy collisions}},}\ }\href {\doibase
  10.1103/PhysRevC.90.034902} {\bibfield  {journal} {\bibinfo  {journal} {Phys.
  Rev. C}\ }\textbf {\bibinfo {volume} {90}},\ \bibinfo {pages} {034902}
  (\bibinfo {year} {2014})}\BibitemShut {NoStop}%
\bibitem [{\citenamefont {Poskanzer}\ and\ \citenamefont
  {Voloshin}(1998)}]{Poskanzer:1998yz}%
  \BibitemOpen
  \bibfield  {author} {\bibinfo {author} {\bibfnamefont {A.~M.}\ \bibnamefont
  {Poskanzer}}\ and\ \bibinfo {author} {\bibfnamefont {S.~A.}\ \bibnamefont
  {Voloshin}},\ }\bibfield  {title} {\enquote {\bibinfo {title} {{Methods for
  analyzing anisotropic flow in relativistic nuclear collisions}},}\ }\href
  {\doibase 10.1103/PhysRevC.58.1671} {\bibfield  {journal} {\bibinfo
  {journal} {Phys. Rev. C}\ }\textbf {\bibinfo {volume} {58}},\ \bibinfo
  {pages} {1671} (\bibinfo {year} {1998})}\BibitemShut {NoStop}%
\bibitem [{\citenamefont {Habich}\ \emph {et~al.}(2015)\citenamefont {Habich},
  \citenamefont {Nagle},\ and\ \citenamefont {Romatschke}}]{Habich:2014jna}%
  \BibitemOpen
  \bibfield  {author} {\bibinfo {author} {\bibfnamefont {M.}~\bibnamefont
  {Habich}}, \bibinfo {author} {\bibfnamefont {J.~L.}\ \bibnamefont {Nagle}}, \
  and\ \bibinfo {author} {\bibfnamefont {P.}~\bibnamefont {Romatschke}},\
  }\bibfield  {title} {\enquote {\bibinfo {title} {{Particle spectra and HBT
  radii for simulated central nuclear collisions of C + C, Al + Al, Cu + Cu, Au
  + Au, and Pb + Pb from $\sqrt{s}=62.4$- $2760$ GeV}},}\ }\href {\doibase
  10.1140/epjc/s10052-014-3206-7} {\bibfield  {journal} {\bibinfo  {journal}
  {Eur. Phys. J. C}\ }\textbf {\bibinfo {volume} {75}},\ \bibinfo {pages} {15}
  (\bibinfo {year} {2015})}\BibitemShut {NoStop}%
\bibitem [{\citenamefont {Adler}\ \emph {et~al.}(2001)\citenamefont {Adler}
  \emph {et~al.}}]{PhysRevLett.87.182301}%
  \BibitemOpen
  \bibfield  {author} {\bibinfo {author} {\bibfnamefont {C.}~\bibnamefont
  {Adler}} \emph {et~al.} (\bibinfo {collaboration} {STAR Collaboration}),\
  }\bibfield  {title} {\enquote {\bibinfo {title} {{Identified particle
  elliptic flow in Au$+$Au collisions at $\sqrt{s_{NN}}=200$ GeV}},}\ }\href
  {\doibase 10.1103/PhysRevLett.87.182301} {\bibfield  {journal} {\bibinfo
  {journal} {Phys. Rev. Lett.}\ }\textbf {\bibinfo {volume} {87}},\ \bibinfo
  {pages} {182301} (\bibinfo {year} {2001})}\BibitemShut {NoStop}%
\bibitem [{\citenamefont {Novak}\ \emph
  {et~al.}(2014{\natexlab{a}})\citenamefont {Novak}, \citenamefont {Novak},
  \citenamefont {Pratt}, \citenamefont {Vredevoogd}, \citenamefont
  {Coleman-Smith},\ and\ \citenamefont {Wolpert}}]{PhysRevC.89.034917}%
  \BibitemOpen
  \bibfield  {author} {\bibinfo {author} {\bibfnamefont {J.}~\bibnamefont
  {Novak}}, \bibinfo {author} {\bibfnamefont {K.}~\bibnamefont {Novak}},
  \bibinfo {author} {\bibfnamefont {S.}~\bibnamefont {Pratt}}, \bibinfo
  {author} {\bibfnamefont {J.}~\bibnamefont {Vredevoogd}}, \bibinfo {author}
  {\bibfnamefont {C.~E.}\ \bibnamefont {Coleman-Smith}}, \ and\ \bibinfo
  {author} {\bibfnamefont {R.~L.}\ \bibnamefont {Wolpert}},\ }\bibfield
  {title} {\enquote {\bibinfo {title} {Determining fundamental properties of
  matter created in ultrarelativistic heavy-ion collisions},}\ }\href {\doibase
  10.1103/PhysRevC.89.034917} {\bibfield  {journal} {\bibinfo  {journal} {Phys.
  Rev. C}\ }\textbf {\bibinfo {volume} {89}},\ \bibinfo {pages} {034917}
  (\bibinfo {year} {2014}{\natexlab{a}})}\BibitemShut {NoStop}%
\bibitem [{\citenamefont {van~der Schee}\ \emph {et~al.}(2013)\citenamefont
  {van~der Schee}, \citenamefont {Romatschke},\ and\ \citenamefont
  {Pratt}}]{vanderSchee:2013pia}%
  \BibitemOpen
  \bibfield  {author} {\bibinfo {author} {\bibfnamefont {W.}~\bibnamefont
  {van~der Schee}}, \bibinfo {author} {\bibfnamefont {P.}~\bibnamefont
  {Romatschke}}, \ and\ \bibinfo {author} {\bibfnamefont {S.}~\bibnamefont
  {Pratt}},\ }\bibfield  {title} {\enquote {\bibinfo {title} {{Fully Dynamical
  Simulation of Central Nuclear Collisions}},}\ }\href {\doibase
  10.1103/PhysRevLett.111.222302} {\bibfield  {journal} {\bibinfo  {journal}
  {Phys. Rev. Lett.}\ }\textbf {\bibinfo {volume} {111}},\ \bibinfo {pages}
  {222302} (\bibinfo {year} {2013})}\BibitemShut {NoStop}%
\bibitem [{\citenamefont {Chesler}\ and\ \citenamefont
  {Yaffe}()}]{Chesler:2015wra}%
  \BibitemOpen
  \bibfield  {author} {\bibinfo {author} {\bibfnamefont {P.~M.}\ \bibnamefont
  {Chesler}}\ and\ \bibinfo {author} {\bibfnamefont {L.~G.}\ \bibnamefont
  {Yaffe}},\ }\href@noop {} {\enquote {\bibinfo {title} {{Holography and
  off-center collisions of localized shock waves}},}\ }\bibinfo {note} {{J.
  High Energy Phys. {\bf 10 (2015)} 070}}\BibitemShut {NoStop}%
\bibitem [{\citenamefont {Romatschke}\ and\ \citenamefont
  {Hogg}()}]{Romatschke:2013re}%
  \BibitemOpen
  \bibfield  {author} {\bibinfo {author} {\bibfnamefont {P.}~\bibnamefont
  {Romatschke}}\ and\ \bibinfo {author} {\bibfnamefont {J.~D.}\ \bibnamefont
  {Hogg}},\ }\href@noop {} {\enquote {\bibinfo {title} {{Pre-Equilibrium Radial
  Flow from Central Shock-Wave Collisions in AdS5}},}\ }\bibinfo {note} {{J.
  High Energy Phys. {\bf 04 (2013)} 048}}\BibitemShut {NoStop}%
\bibitem [{\citenamefont {Adare}\ \emph
  {et~al.}(2016{\natexlab{b}})\citenamefont {Adare} \emph
  {et~al.}}]{Adare:2015bua}%
  \BibitemOpen
  \bibfield  {author} {\bibinfo {author} {\bibfnamefont {A.}~\bibnamefont
  {Adare}} \emph {et~al.} (\bibinfo {collaboration} {{PHENIX Collaboration}}),\
  }\bibfield  {title} {\enquote {\bibinfo {title} {{Transverse energy
  production and charged-particle multiplicity at midrapidity in various
  systems from $\sqrt{s_{NN}}=7.7$ to 200 GeV}},}\ }\href {\doibase
  10.1103/PhysRevC.93.024901} {\bibfield  {journal} {\bibinfo  {journal} {Phys.
  Rev. C}\ }\textbf {\bibinfo {volume} {93}},\ \bibinfo {pages} {024901}
  (\bibinfo {year} {2016}{\natexlab{b}})}\BibitemShut {NoStop}%
\bibitem [{\citenamefont {Huovinen}\ \emph {et~al.}(2001)\citenamefont
  {Huovinen}, \citenamefont {Kolb}, \citenamefont {Heinz}, \citenamefont
  {Ruuskanen},\ and\ \citenamefont {Voloshin}}]{Huovinen:2001cy}%
  \BibitemOpen
  \bibfield  {author} {\bibinfo {author} {\bibfnamefont {P.}~\bibnamefont
  {Huovinen}}, \bibinfo {author} {\bibfnamefont {P.~F.}\ \bibnamefont {Kolb}},
  \bibinfo {author} {\bibfnamefont {U.~W.}\ \bibnamefont {Heinz}}, \bibinfo
  {author} {\bibfnamefont {P.~V.}\ \bibnamefont {Ruuskanen}}, \ and\ \bibinfo
  {author} {\bibfnamefont {S.~A.}\ \bibnamefont {Voloshin}},\ }\bibfield
  {title} {\enquote {\bibinfo {title} {{Radial and elliptic flow at RHIC:
  Further predictions}},}\ }\href {\doibase 10.1016/S0370-2693(01)00219-2}
  {\bibfield  {journal} {\bibinfo  {journal} {Phys. Lett. B}\ }\textbf
  {\bibinfo {volume} {503}},\ \bibinfo {pages} {58} (\bibinfo {year}
  {2001})}\BibitemShut {NoStop}%
\bibitem [{\citenamefont {Fries}\ \emph {et~al.}(2008)\citenamefont {Fries},
  \citenamefont {Greco},\ and\ \citenamefont {Sorensen}}]{Fries:2008hs}%
  \BibitemOpen
  \bibfield  {author} {\bibinfo {author} {\bibfnamefont {R.~J.}\ \bibnamefont
  {Fries}}, \bibinfo {author} {\bibfnamefont {V.}~\bibnamefont {Greco}}, \ and\
  \bibinfo {author} {\bibfnamefont {P.}~\bibnamefont {Sorensen}},\ }\bibfield
  {title} {\enquote {\bibinfo {title} {{Coalescence Models For Hadron Formation
  From Quark Gluon Plasma}},}\ }\href {\doibase
  10.1146/annurev.nucl.58.110707.171134} {\bibfield  {journal} {\bibinfo
  {journal} {Ann. Rev. Nucl. Part. Sci.}\ }\textbf {\bibinfo {volume} {58}},\
  \bibinfo {pages} {177} (\bibinfo {year} {2008})}\BibitemShut {NoStop}%
\bibitem [{\citenamefont {Adler}\ \emph
  {et~al.}(2003{\natexlab{b}})\citenamefont {Adler} \emph
  {et~al.}}]{Adler:2003kg}%
  \BibitemOpen
  \bibfield  {author} {\bibinfo {author} {\bibfnamefont {S.~S.}\ \bibnamefont
  {Adler}} \emph {et~al.} (\bibinfo {collaboration} {{PHENIX Collaboration}}),\
  }\bibfield  {title} {\enquote {\bibinfo {title} {{Scaling properties of
  proton and anti-proton production in $\sqrt{s_{NN}}=200$ GeV Au+Au
  collisions}},}\ }\href {\doibase 10.1103/PhysRevLett.91.172301} {\bibfield
  {journal} {\bibinfo  {journal} {Phys. Rev. Lett.}\ }\textbf {\bibinfo
  {volume} {91}},\ \bibinfo {pages} {172301} (\bibinfo {year}
  {2003}{\natexlab{b}})}\BibitemShut {NoStop}%
\bibitem [{\citenamefont {Adler}\ \emph {et~al.}(2004)\citenamefont {Adler}
  \emph {et~al.}}]{Adler:2003cb}%
  \BibitemOpen
  \bibfield  {author} {\bibinfo {author} {\bibfnamefont {S.~S.}\ \bibnamefont
  {Adler}} \emph {et~al.} (\bibinfo {collaboration} {{PHENIX Collaboration}}),\
  }\bibfield  {title} {\enquote {\bibinfo {title} {{Identified charged particle
  spectra and yields in Au+Au collisions at $\sqrt{s_{NN}}=200$ GeV}},}\ }\href
  {\doibase 10.1103/PhysRevC.69.034909} {\bibfield  {journal} {\bibinfo
  {journal} {Phys. Rev. C}\ }\textbf {\bibinfo {volume} {69}},\ \bibinfo
  {pages} {034909} (\bibinfo {year} {2004})}\BibitemShut {NoStop}%
\bibitem [{\citenamefont {Bass}\ \emph {et~al.}(1998)\citenamefont {Bass} \emph
  {et~al.}}]{Bass:1998ca}%
  \BibitemOpen
  \bibfield  {author} {\bibinfo {author} {\bibfnamefont {S.~A.}\ \bibnamefont
  {Bass}} \emph {et~al.},\ }\bibfield  {title} {\enquote {\bibinfo {title}
  {{Microscopic models for ultrarelativistic heavy ion collisions}},}\ }\href
  {\doibase 10.1016/S0146-6410(98)00058-1} {\bibfield  {journal} {\bibinfo
  {journal} {Prog. Part. Nucl. Phys.}\ }\textbf {\bibinfo {volume} {41}},\
  \bibinfo {pages} {255} (\bibinfo {year} {1998})}\BibitemShut {NoStop}%
\bibitem [{\citenamefont {Bleicher}\ \emph {et~al.}(1999)\citenamefont
  {Bleicher} \emph {et~al.}}]{Bleicher:1999xi}%
  \BibitemOpen
  \bibfield  {author} {\bibinfo {author} {\bibfnamefont {M.}~\bibnamefont
  {Bleicher}} \emph {et~al.},\ }\bibfield  {title} {\enquote {\bibinfo {title}
  {{Relativistic hadron hadron collisions in the ultrarelativistic quantum
  molecular dynamics model}},}\ }\href {\doibase 10.1088/0954-3899/25/9/308}
  {\bibfield  {journal} {\bibinfo  {journal} {J. Phys. G}\ }\textbf {\bibinfo
  {volume} {25}},\ \bibinfo {pages} {1859} (\bibinfo {year}
  {1999})}\BibitemShut {NoStop}%
\bibitem [{\citenamefont {Adare}\ \emph
  {et~al.}(2016{\natexlab{c}})\citenamefont {Adare} \emph
  {et~al.}}]{Adare:2015cpn}%
  \BibitemOpen
  \bibfield  {author} {\bibinfo {author} {\bibfnamefont {A.}~\bibnamefont
  {Adare}} \emph {et~al.} (\bibinfo {collaboration} {{PHENIX Collaboration}}),\
  }\bibfield  {title} {\enquote {\bibinfo {title} {{Measurements of directed,
  elliptic, and triangular flow in Cu$+$Au collisions at $\sqrt{s_{_{NN}}}=200$
  GeV}},}\ }\href {\doibase 10.1103/PhysRevC.94.054910} {\bibfield  {journal}
  {\bibinfo  {journal} {Phys. Rev. C}\ }\textbf {\bibinfo {volume} {94}},\
  \bibinfo {pages} {054910} (\bibinfo {year} {2016}{\natexlab{c}})}\BibitemShut
  {NoStop}%
\bibitem [{\citenamefont {Orjuela~Koop}\ \emph {et~al.}(2015)\citenamefont
  {Orjuela~Koop}, \citenamefont {Adare}, \citenamefont {McGlinchey},\ and\
  \citenamefont {Nagle}}]{Koop:2015wea}%
  \BibitemOpen
  \bibfield  {author} {\bibinfo {author} {\bibfnamefont {J.~D.}\ \bibnamefont
  {Orjuela~Koop}}, \bibinfo {author} {\bibfnamefont {A.}~\bibnamefont {Adare}},
  \bibinfo {author} {\bibfnamefont {D.}~\bibnamefont {McGlinchey}}, \ and\
  \bibinfo {author} {\bibfnamefont {J.~L.}\ \bibnamefont {Nagle}},\ }\bibfield
  {title} {\enquote {\bibinfo {title} {{Azimuthal anisotropy relative to the
  participant plane from a multiphase transport model in central p + Au , d +
  Au , and $^{3}$He + Au collisions at $\sqrt{s_{NN}}=200$ GeV}},}\ }\href
  {\doibase 10.1103/PhysRevC.92.054903} {\bibfield  {journal} {\bibinfo
  {journal} {Phys. Rev. C}\ }\textbf {\bibinfo {volume} {92}},\ \bibinfo
  {pages} {054903} (\bibinfo {year} {2015})}\BibitemShut {NoStop}%
\bibitem [{\citenamefont {Orjuela~Koop}\ \emph {et~al.}(2016)\citenamefont
  {Orjuela~Koop}, \citenamefont {Belmont}, \citenamefont {Yin},\ and\
  \citenamefont {Nagle}}]{Koop:2015trj}%
  \BibitemOpen
  \bibfield  {author} {\bibinfo {author} {\bibfnamefont {J.~D.}\ \bibnamefont
  {Orjuela~Koop}}, \bibinfo {author} {\bibfnamefont {R.}~\bibnamefont
  {Belmont}}, \bibinfo {author} {\bibfnamefont {P.}~\bibnamefont {Yin}}, \ and\
  \bibinfo {author} {\bibfnamefont {J.~L.}\ \bibnamefont {Nagle}},\ }\bibfield
  {title} {\enquote {\bibinfo {title} {{Exploring the Beam Energy Dependence of
  Flow-Like Signatures in Small System $d+$Au Collisions}},}\ }\href {\doibase
  10.1103/PhysRevC.93.044910} {\bibfield  {journal} {\bibinfo  {journal} {Phys.
  Rev. C}\ }\textbf {\bibinfo {volume} {93}},\ \bibinfo {pages} {044910}
  (\bibinfo {year} {2016})}\BibitemShut {NoStop}%
\bibitem [{\citenamefont {Ma}\ and\ \citenamefont {Lin}(2016)}]{Ma:2016fve}%
  \BibitemOpen
  \bibfield  {author} {\bibinfo {author} {\bibfnamefont {G.~L.}\ \bibnamefont
  {Ma}}\ and\ \bibinfo {author} {\bibfnamefont {Z.~W.}\ \bibnamefont {Lin}},\
  }\bibfield  {title} {\enquote {\bibinfo {title} {{Predictions for $\sqrt
  {s_{NN}}=5.02$ TeV Pb$+$Pb Collisions from a Multi-Phase Transport Model}},}\
  }\href {\doibase 10.1103/PhysRevC.93.054911} {\bibfield  {journal} {\bibinfo
  {journal} {Phys. Rev. C}\ }\textbf {\bibinfo {volume} {93}},\ \bibinfo
  {pages} {054911} (\bibinfo {year} {2016})}\BibitemShut {NoStop}%
\bibitem [{\citenamefont {Ma}\ and\ \citenamefont
  {Bzdak}(2014)}]{ma_long-range_2014}%
  \BibitemOpen
  \bibfield  {author} {\bibinfo {author} {\bibfnamefont {G.}~\bibnamefont
  {Ma}}\ and\ \bibinfo {author} {\bibfnamefont {A.}~\bibnamefont {Bzdak}},\
  }\bibfield  {title} {\enquote {\bibinfo {title} {{Long-range azimuthal
  correlations in proton-proton and proton-nucleus collisions from the
  incoherent scattering of partons}},}\ }\href {\doibase
  http://dx.doi.org/10.1016/j.physletb.2014.10.066} {\bibfield  {journal}
  {\bibinfo  {journal} {Phys. Lett. B}\ }\textbf {\bibinfo {volume} {739}},\
  \bibinfo {pages} {209} (\bibinfo {year} {2014})}\BibitemShut {NoStop}%
\bibitem [{\citenamefont {Li}\ and\ \citenamefont
  {Ko}(1995)}]{PhysRevC.52.2037}%
  \BibitemOpen
  \bibfield  {author} {\bibinfo {author} {\bibfnamefont {B.~A.}\ \bibnamefont
  {Li}}\ and\ \bibinfo {author} {\bibfnamefont {C.~M.}\ \bibnamefont {Ko}},\
  }\bibfield  {title} {\enquote {\bibinfo {title} {Formation of superdense
  hadronic matter in high energy heavy-ion collisions},}\ }\href {\doibase
  10.1103/PhysRevC.52.2037} {\bibfield  {journal} {\bibinfo  {journal} {Phys.
  Rev. C}\ }\textbf {\bibinfo {volume} {52}},\ \bibinfo {pages} {2037}
  (\bibinfo {year} {1995})}\BibitemShut {NoStop}%
\bibitem [{\citenamefont {Novak}\ \emph
  {et~al.}(2014{\natexlab{b}})\citenamefont {Novak}, \citenamefont {Novak},
  \citenamefont {Pratt}, \citenamefont {Vredevoogd}, \citenamefont
  {Coleman-Smith},\ and\ \citenamefont {Wolpert}}]{Novak:2013bqa}%
  \BibitemOpen
  \bibfield  {author} {\bibinfo {author} {\bibfnamefont {J.}~\bibnamefont
  {Novak}}, \bibinfo {author} {\bibfnamefont {K.}~\bibnamefont {Novak}},
  \bibinfo {author} {\bibfnamefont {S.}~\bibnamefont {Pratt}}, \bibinfo
  {author} {\bibfnamefont {J.}~\bibnamefont {Vredevoogd}}, \bibinfo {author}
  {\bibfnamefont {C.~E.}\ \bibnamefont {Coleman-Smith}}, \ and\ \bibinfo
  {author} {\bibfnamefont {R.~L.}\ \bibnamefont {Wolpert}},\ }\bibfield
  {title} {\enquote {\bibinfo {title} {{Determining Fundamental Properties of
  Matter Created in Ultrarelativistic Heavy-Ion Collisions}},}\ }\href
  {\doibase 10.1103/PhysRevC.89.034917} {\bibfield  {journal} {\bibinfo
  {journal} {Phys. Rev. C}\ }\textbf {\bibinfo {volume} {89}},\ \bibinfo
  {pages} {034917} (\bibinfo {year} {2014}{\natexlab{b}})}\BibitemShut
  {NoStop}%
\bibitem [{\citenamefont {Adare}\ \emph {et~al.}(2007)\citenamefont {Adare}
  \emph {et~al.}}]{Adare:2006ti}%
  \BibitemOpen
  \bibfield  {author} {\bibinfo {author} {\bibfnamefont {A.}~\bibnamefont
  {Adare}} \emph {et~al.} (\bibinfo {collaboration} {{PHENIX Collaboration}}),\
  }\bibfield  {title} {\enquote {\bibinfo {title} {{Scaling properties of
  azimuthal anisotropy in Au+Au and Cu+Cu collisions at s(NN) = 200-GeV}},}\
  }\href {\doibase 10.1103/PhysRevLett.98.162301} {\bibfield  {journal}
  {\bibinfo  {journal} {Phys. Rev. Lett.}\ }\textbf {\bibinfo {volume} {98}},\
  \bibinfo {pages} {162301} (\bibinfo {year} {2007})}\BibitemShut {NoStop}%
\bibitem [{\citenamefont {Adams}\ \emph {et~al.}(2004)\citenamefont {Adams}
  \emph {et~al.}}]{PhysRevLett.92.052302}%
  \BibitemOpen
  \bibfield  {author} {\bibinfo {author} {\bibfnamefont {John}\ \bibnamefont
  {Adams}} \emph {et~al.} (\bibinfo {collaboration} {STAR Collaboration}),\
  }\bibfield  {title} {\enquote {\bibinfo {title} {{Particle type dependence of
  azimuthal anisotropy and nuclear modification of particle production in
  Au$+$Au collisions at $\sqrt{s_{NN}}=200$ GeV}},}\ }\href {\doibase
  10.1103/PhysRevLett.92.052302} {\bibfield  {journal} {\bibinfo  {journal}
  {Phys. Rev. Lett.}\ }\textbf {\bibinfo {volume} {92}},\ \bibinfo {pages}
  {052302} (\bibinfo {year} {2004})}\BibitemShut {NoStop}%
\bibitem [{\citenamefont {Adare}\ \emph {et~al.}(2012)\citenamefont {Adare}
  \emph {et~al.}}]{Adare:2012vq}%
  \BibitemOpen
  \bibfield  {author} {\bibinfo {author} {\bibfnamefont {A.}~\bibnamefont
  {Adare}} \emph {et~al.} (\bibinfo {collaboration} {{PHENIX Collaboration}}),\
  }\bibfield  {title} {\enquote {\bibinfo {title} {{Deviation from quark-number
  scaling of the anisotropy parameter $v_2$ of pions, kaons, and protons in
  Au+Au collisions at $\sqrt{s_{NN}} = 200$ GeV}},}\ }\href {\doibase
  10.1103/PhysRevC.85.064914} {\bibfield  {journal} {\bibinfo  {journal} {Phys.
  Rev. C}\ }\textbf {\bibinfo {volume} {85}},\ \bibinfo {pages} {064914}
  (\bibinfo {year} {2012})}\BibitemShut {NoStop}%
\end{thebibliography}

%
 
\end{document}